\documentclass[a4paper,fleqn,usenatbib]{mnras}
\usepackage[T1]{fontenc}
\usepackage{ae,aecompl}

\usepackage{graphicx}% Including figure files
\usepackage{amsmath}% Advanced maths commands
\usepackage{amssymb}% Extra maths symbols

\usepackage[lofdepth,lotdepth,caption=false]{subfig}% for subfloats

\title[Exploring a model for quasi-periodic light curves]{Variable classification in the LSST era: Exploring a model for quasi-periodic light curves}

\author[Zinn et al.]{
J.~C. Zinn,$^{1}$\thanks{E-mail: zinn.44@osu.edu (JCZ)}
C.~S. Kochanek,$^{1,2}$
S. Koz{\l}owski,$^3$
A. Udalski,$^3$
M.~K. Szyma{\'n}ski,$^3$
\newauthor
I. Soszy{\'n}ski,$^3$
{\L}. Wyrzykowski,$^3$
K. Ulaczyk,$^{3,4}$
R. Poleski,$^{3,1}$
P. Pietrukowicz,$^3$
\newauthor
J. Skowron,$^3$
P. Mr{\'o}z,$^3$
and M. Pawlak$^3$
\\
$^{1}$ Department of Astronomy, The Ohio State University, 140 West 18th Avenue, Columbus OH 43210, USA\\ 
$^{2}$ Center for Cosmology and AstroParticle Physics, The Ohio State University, 191 W. Woodruff Avenue, Columbus OH 43210, USA\\ 
$^3$ Warsaw University Observatory, Al.~Ujazdowskie~4, 00-478~Warszawa, Poland\\
$^4$ Department of Physics, University of Warwick, Gibbet Hill Road, Coventry, CV4~7AL,~UK
} 

\date{Accepted XXX. Received YYY; in original form ZZZ}

\pubyear{2016}

\begin{document}
\label{firstpage}
\pagerange{\pageref{firstpage}--\pageref{lastpage}}
\maketitle

\begin{abstract} 
LSST is expected to yield $\sim 10^7$ light curves over the course of
its mission, which will require a concerted effort in automated
classification. Stochastic processes provide one means of
quantitatively describing variability with the potential advantage
over simple light curve statistics that the parameters may be
physically meaningful. Here, we survey a
large sample of periodic, quasi-periodic, and stochastic OGLE-III
variables using the damped random walk (DRW; CARMA(1,0)) and quasi-periodic
oscillation (QPO; CARMA(2,1)) stochastic process models. The QPO model
is described by an amplitude, a period, and a coherence time-scale,
while the DRW has only an amplitude and a time-scale. We
find that the periodic and quasi-periodic stellar variables are
generally better
described by a QPO than a DRW, while quasars are better described by
the DRW model. There are ambiguities in interpreting
the QPO coherence time due to non-sinusoidal
light curve shapes, signal-to-noise, error mischaracterizations, and
cadence. Higher-order implementations of the QPO model that better capture
light curve shapes are necessary for the coherence time to have its
implied physical meaning. Independent of physical meaning, the extra
parameter of the QPO model successfully distinguishes most of the
classes of periodic and quasi-periodic variables we consider.
\end{abstract} 
 
 
\section{Introduction} 
\label{sec:intro} 
Time-domain astronomy has resulted in a range of observed
astrophysical variability regimes that may be broadly categorized as
periodic, quasi-periodic, stochastic, and transient. Transient events
are short time-scale changes in flux such as gamma ray bursts \citep{klebesadel1973,cano+2016},
supernovae \citep[for a review see, e.g.,][]{woosley_weaver1995}, cataclysmic variables \citep{robinson1976}, and stellar
flares \citep[Variable Star Network;][]{kato+2004}. These can
frequently be modeled with a template light curve whose structure and
parameters can be used to classify the transient \citep[e.g., photometric SN
  classification,][]{pskovskii1977,hamuy+1996,sako+2008,sako+2011}. Periodic
variables such as RR Lyrae, Cepheids, and eclipsing binaries are
classified based on their period and the structure of their phased
light curves
\citep[e.g.,][]{pojmanski2002,debosscher2007,soszynski+2008,sarro+2009,soszynski+2009a,graczyk+2011}. Quasi-periodic
sources such as Miras or spotted stars have one or more dominant
frequencies describing their variability, but do not maintain
consistent phase and/or amplitude as a function of time
\citep[e.g.][]{howarth1991,bedding+2005}. In this regime, sources are
classified by a period and average light curve structure \citep[usually
Fourier component ratios, e.g.,][]{soszynski+2008}, but there is no measure of the coherence of the
variability. Stochastic variables such as AGN vary without any obvious
pattern and may be modeled by stochastic processes \citep[e.g.,][]{kelly_bechtold_siemiginowska2009, kozlowski+2010, macleod+2010, zu+2013}.
 
While studies of periodic sources are ubiquitous in astronomy, there
are far fewer quantitative studies of quasi-periodic
systems. Presently, the primary large-scale application is the
determination of stellar rotation rates for \textit{Kepler} stars
\citep{balaji+2015,mcquillan_aigrain_mazeh2013,reinhold+2013}. In this case, the variability is
quasi-periodic due to the finite lifetime of star spots and
differential rotation. These periods have been used to study rotation
rates in stars \citep[e.g.,][]{frasca+2011}, and are the basis for
gyrochronology, where stellar ages are estimated based on models for
spin-down rates \citep{garcia+2014}. Quasi-periodic variables such as
Long-Period variables (LPVs) are also well known. Qualitative measures
of irregularity in light curves have been used for classification,
most notably in the General Catalogue of Variable stars
\citep{kholopov+1985} and the Optical Gravitational Lensing Experiment
Catalogue of Variable Stars \citep{soszynski+2008}, but quantification of the degree of
quasi-periodicity for these sources remains little-explored.

The study of variable sources benefits from a broad range of
surveys. There are many ongoing surveys for transients such as the
Optical Gravitational Lensing Experiment \citep[OGLE;
  e.g.,][]{udalski+2008}, the Catalina Real-Time Transient Survey
\citep{drake+2009}, the Palomar Transient Factory \citep{law+2009},
QUEST \citep{snyder1998}, the All-Sky Automated Survey
\citep{pojmanski1997}, and the All-Sky Automated Survey for SuperNovae
\citep{shappee+2014}. Apart from these variability surveys, searches
for transiting exoplanets such as the Hungarian Automated Telescope
\citep[HAT,][]{bakos+2002}, \textit{Kepler} \citep{borucki+2010}, CoRoT
\citep{auvergne+2009}, KELT \citep{pepper_gould_depoy2004}, and
SuperWASP \citep{pollacco+2006,smith+2014}  have also produced a
wealth of light curves for other variable phenomena. More general
surveys such as the Sloan Digital Sky Survey \citep{york+2000},
Pan-STARRS \citep{kaiser+2002}, and the Dark Energy Survey
\citep{DES2005} also include time domain components. Future surveys
such as The Large Synaptic Survey Telescope \citep[LSST;][]{ivezic+2008, tyson2002, lsst2009}, TESS \citep{ricker+2014}, PLATO 2.0, and the Zwicky Transient Facility \citep{smith+2014} promise to expand time domain studies still further.
 
Two challenges to science with variability surveys are to first
classify the variable sources and then to extract physical information
about the variability from the light curves. A promising avenue for
parsing the huge numbers of light curves is machine learning
(e.g., \citealt{wozniak+2004,debosscher2007, richards+2012,bloom+2012};
see \citealt{kim+2014,ball&brunner2010} for reviews of machine learning in astronomy). In most
machine learning approaches, a large number of statistics are
calculated for each light curve and an algorithm is trained to
classify variables based on the properties of a sample of previously
classified variables \citep[but see, e.g.,][for an example of
  unsupervised classification of variable stars]{mackenzie2016}. While these methods work well for identifying
transients such as supernovae and highly periodic sources like
Cepheids and RR Lyrae, stars with irregular periodicity and
multi-periodicity have proven more difficult to accurately classify. One problem
may be that the statistics employed in many of the current approaches
have no physical significance. Parameters with physical meaning such
as periods may be better for classifying light curves than {\it ad
  hoc} statistical measures, in addition to providing insights into
the nature of the variable source. The challenge is the definition and
extraction of new physically meaningful parameters from quasi-periodic
or stochastic light curves.

The introduction of stochastic process models for quasar variability
\citep{kelly_bechtold_siemiginowska2009, kozlowski+2010} is a good
recent demonstration of these principles. A damped random walk (DRW)
stochastic process characterized by an amplitude and a coherence
time-scale provides a good, compact statistical model of quasar
variables on time-scales of days to decades \citep{macleod+2012}.  The
parameters of the model are then correlated with the observed
wavelength and the intrinsic properties of the quasars
\citep[e.g.,][]{macleod+2010}, potentially providing insight into the
physical origin of the variability
\citep{kelly_bechtold_siemiginowska2009, dexter&agol2011,
  andrae_kim_bailer-jones2013}. There is evidence for deviations from
the DRW, particularly on short time-scales
\citep[e.g.,][]{mushotzky+2011, zu+2013, graham+2014,
  andrae_kim_bailer-jones2013, kasliwal_vogeley_richards2015}, and
this may allow the extraction of additional physical scales from AGN
light curves. The DRW model has also seen application in
\cite{he+2016} to model Miras, in which they find that adding a DRW
component in addition to a purely
periodic component can improve the fidelity of period estimates by up
to $20$ per cent.
 
The DRW model is the first ($p = 1$) in a series of continuous time
autoregressive moving average (CARMA($p,r$)) processes, which are the
solutions to stochastic linear differential equations of order $p$
driven by white noise and up to $r < p$ derivatives of white noise
processes. The higher order processes in this sequence have
occasionally been used in astronomy \citep[e.g.,][]{koen2005,
  koen2012, andrae_kim_bailer-jones2013, kelly+2014,
  aigrain_parviainen_pope2016, kasliwal_vogeley_richards2015}. In
particular, \cite{kelly+2014} offer a package implementing fits of
these models to light curves based on methods from the forecasting
literature \citep[e.g.,][]{brockwell&davis2002}. 

Physically, the $p = 2$ process models a quasi-periodic oscillation
(QPO) characterized by an amplitude, a period, and a coherence
time. \cite{koen2005, koen2012}, \cite{kelly+2014}, and
\cite{aigrain_parviainen_pope2016} give examples of applying the model
to periodic and quasi-periodic variables, but it has never been
employed for a large-scale variability survey. Here we carry out such
a survey, applying the QPO model to large numbers of variable stars
and quasars in the OGLE-III fields for the Large Magellanic Cloud
(LMC). We also compare the QPO results to those from modeling the same
sources using a DRW, and explore to what extent certain variables
prefer a QPO to traditional purely periodic models. The goals are to
explore whether the two models successfully distinguish between
variable stars and quasars and whether the QPO coherence time provides
a useful means of classifying and understanding variable stars. In
\S\ref{sec:data} we describe the light curve data and the model
formalism. In \S\ref{sec:understanding} we provide a guide towards
understanding the results presented in \S\ref{sec:results}, and we conclude in \S\ref{sec:conclusion} with a discussion.
 
\section{Data and Methods} 
\label{sec:data} 
The Optical Gravitational Lensing Experiment (OGLE) observes the
Magellanic Clouds, and portions of the Galactic bulge and disk in the
I- and V-bands using the 1.3-m Warsaw telescope at the Las Campanas
Observatory. We use light curves of sources in or behind the LMC from
the OGLE-III survey \citep{udalski+2008}, which operated from 2001 to
2009. We examine 3361 classical Cepheids \citep{soszynski+2008}, a
subset of 50000
Long Period Variables \citep{soszynski+2009b},
and 24906 RR Lyrae \citep{soszynski+2009a}, as well
as well as OGLE-III/IV light curves of 753 QSOs identified in the Magellanic Quasars Survey
\citep[MQS,][]{kozlowski+2013}. We used only the
I-band data as they have a better mean cadence ($\sim 6$~days) over a
total baseline of $\sim 3000$~days. We removed MQS QSOs fainter than
$I = 19.5$~mag from our analysis due to degraded light curve quality. We require that each object
strongly prefer (at $95$ per cent confidence) either a DRW or QPO model above pure
white noise, leaving 1460 Cepheids, 24542 Long Period Variables, 24223
RR Lyrae, and 443 QSOs (details below). This requirement ensures that
sufficient correlations can be detected in the light curve to
distinguish a stochastic process model from a simple broadening of
the photometric uncertainties (a ``white noise'' model). The errors on the
OGLE-III data points are calculated using the difference image analysis
(DIA) software described in \cite{wozniak2000}. In many cases, these
errors may be under-estimated. The errors on the QSOs and Cepheids can be
corrected by rescaling the DIA errors to $\sigma'_i = \sqrt{
  (\gamma \sigma_i)^2 + \epsilon^2}$, to match the observed dispersion in standard stars as a
function of magnitude, where $\gamma$ and $\epsilon$
are computed on a field-to-field basis as described
in \citep{wyrzykowski+2009}.

The OGLE LPVs have been broadly categorized based on light curve
properties into Semi-Regular Variables, Miras, and OGLE
Small-Amplitude Semi-Regular Variables (OSARGs) by \cite{soszynski+2009b}.  All three types are understood to be giant branch or asymptotic giant branch stars. OSARGs have variability amplitudes ranging from 0.005 to 0.13 mag with periods ranging from 10 to 100~days \citep{wray_eyer_paczynski2004}, and the pulsation mechanisms responsible for their light curves are not well understood \citep[but see][]{soszynski+2004, christensen-Dalsgaard_kjeldsen_mattei2001}. Miras are generally believed to be variable as a result of inherently unstable radial modes due to opacity effects in the atmosphere \citep[e.g.,][]{wood1974}, with periods ranging from 100 to 1000~days and amplitudes of up to 10 mag. Semi-Regular Variables (SRVs) have similar periods to Miras, but are distinguished from Miras by their lower amplitudes and because they pulsate in both the fundamental and the overtone radial modes, instead of only the fundamental mode. Although their light curves exhibit qualitatively different light curves, with SRVs generally having less coherent periodicity than Miras, \cite{soszynski_wood_udalski2013} hypothesize that as red giants evolve, they first pulsate as OSARGs, then as SRVs, and finally as Miras \citep[see also][]{bedding_zijlstra1998}. 

Among the best-studied variable stars are Cepheid variables---giant
stars with pulsation periods mostly from 2 to 20~days and amplitudes
from 0.2 to 1 mag in V-band due to the $\kappa$-mechanism
\citep{zhevakin1953}. The OGLE Cepheids are divided into sub-classes
based on their pulsation modes, from fundamental mode (``F'') and
first overtone (``1'') pulsators up to third overtone pulsators, with
a small number of multi-mode pulsators \citep{soszynski+2008}. RR
Lyrae stars are horizontal branch stars in the instability strip that
also pulsate due to the $\kappa$-mechanism \citep{christy1966}. Based
on their periods and light curve shapes, RR Lyrae are subdivided into RRab (fundamental mode pulsators), RRc (first overtone pulsators), RRe (second overtone pulsators) and RRd (simultaneous fundamental and first overtone pulsators). Typical variability amplitudes for RR Lyrae range from 0.2 mag to 2 mag in V-band, and the pulsation periods range from of 0.2 to 1~days. 

We model light curves using a family of stochastic process models known as Continuous time Autoregressive Moving Averages (CARMA). These models statistically describe the solution to stochastically-driven ODEs by way of an autocovariance function, $S \equiv \langle \vec{s} | \vec{s}\rangle$, whose structure depends on the order of the ODE, $p$, and the weights given to the $r< p$ noise processes driving the system. For the purposes of this paper, we call the $p = 1$ case a damped random walk (DRW) and the $p = 2$ case a quasi-periodic oscillation (QPO). 
The autocovariance functions of the CARMA processes are the sum of $p$
exponentials $e^{\omega_it}$, where the $\omega_i$ must be real or
appear in complex conjugate pairs \citep[see][]{brockwell_davis1996}. The DRW model is the lowest order example, and its covariance function between times $t_i$ and $t_j$ is
\begin{equation} 
S_{ij} = \sigma_{DRW}^2 \exp{}\left(-|t_j - t_i|/\tau\right), 
\end{equation} 
where $\sigma_{DRW}$ describes the variance of the light curve on long time-scales and $\tau$ is the coherence time. We use the DRW for comparison to the QPO model in order to explore whether a quasi-periodic model is a better model for the stellar variable sources (or the reverse for quasars). The QPO model is the case $p=2$, with a complex conjugate pair of $w_i$, which we can divide into an oscillation period, $P = 2\pi/\omega$, and a coherence time, $\tau$, to give the autocovariance function 
\begin{equation} 
\label{eq:kerq} 
S_{ij} = \frac{\sigma^2}{2} \exp{}\left[-|(t_j - t_i)|/\tau\right]\cos{}\left[\omega(t_j-t_i)\right]. 
\end{equation} 
We could also add a phase, but we omit a phase factor in favour of assuming that all real variable sources have a maximum  
in their autocovariance functions as $|t_j - t_i| \rightarrow 0 $.

\begin{figure*}
\centering
\includegraphics[width=\textwidth]{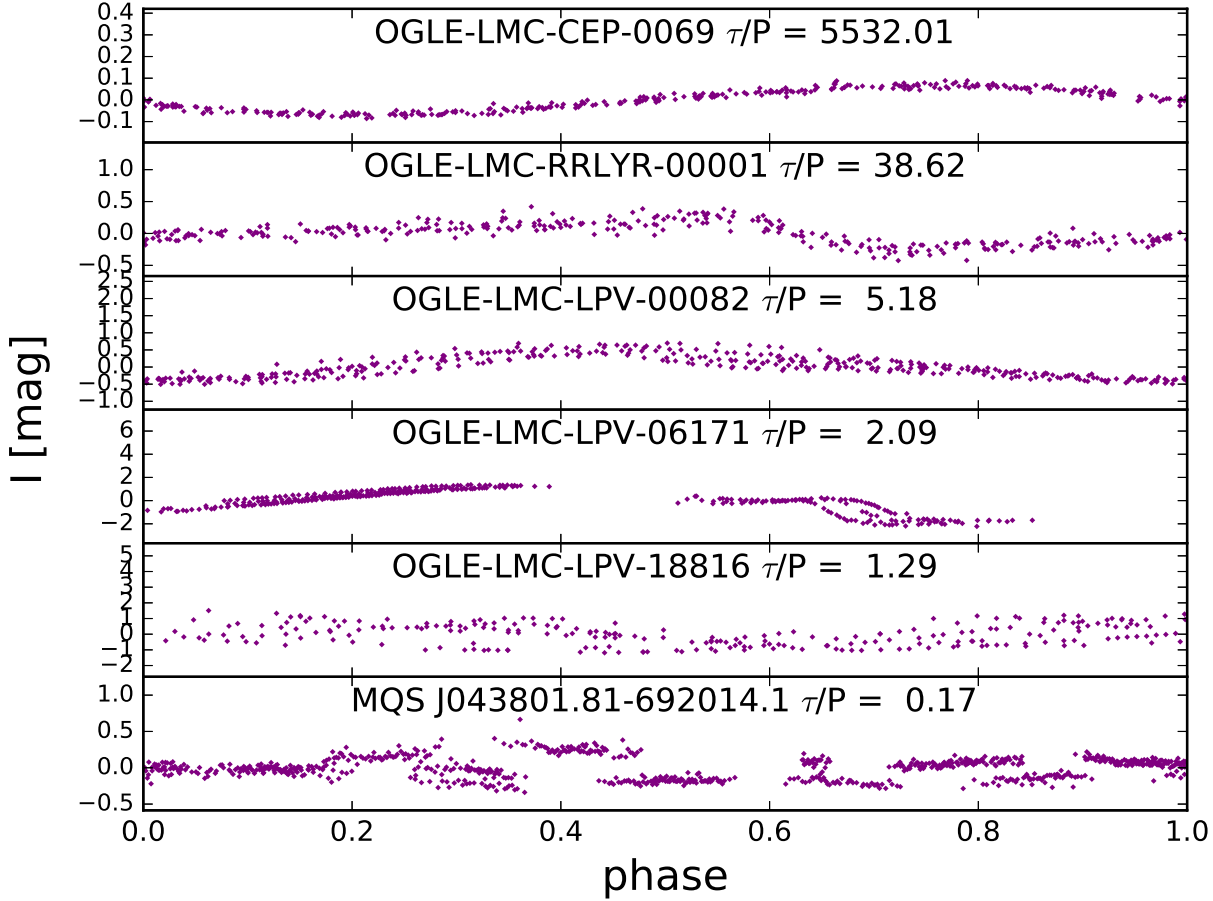}
\caption{Folded light curves illustrating the meaning of the coherence
  time-scale in the QPO model. Each panel shows a folded light curve of
  a source ranging from coherent ($\tau/P\gg 1$) sources at the top to
  incoherent ($\tau/P \ll 1$) sources at the bottom. The most coherent
  source shown is the Cepheid OGLE-LMC-CEP-0069, followed by an RR
  Lyrae, a series of LPVs, and ending with the quasar MQS J043801.81-692014.1. That
$\tau/P$ is not larger for OGLE-LMC-RRLYR-00001 is discussed in \S\protect\ref{sec:understanding}.}
\label{fig:bestfit}
\end{figure*}

Whereas the DRW $\tau$ parameter indicates the timescale over which 
variability is correlated, the QPO $\tau$ parameter is the
time-scale over which the process appears as a coherent
oscillation. Thus, the ratio of the coherence time to the period, $\tau/P$, roughly
corresponds to the number of oscillations over which the light curve
will maintain phase coherence. Figure~\ref{fig:bestfit} illustrates this meaning of the coherence time with a series of phased light curves ranging
from an essentially periodic Cepheid at the top to a non-periodic MQS
quasar at the bottom.  We adopt $\tau/P>10^4$ as
our operational definition of periodic, although RR Lyrae and Cepheid variables can be even more strictly periodic. In practice,
we cannot distinguish among models with $\tau > 10^4$~days due to the OGLE time
baseline \citep[cf., the discussion of Figure~\ref{fig:Abaseline},
below; see also][]{kozlowski2016c}. When fitting models, we impose
$\tau < 10^4$~days.
In the limit that $\tau/P \ll 1$, illustrated by the quasar, the QPO model becomes the DRW model. 

In addition to the process model for the variability, we assume the data have uncorrelated Gaussian errors 
(noise) with an autocovariance function $N \equiv \langle \vec{n} \vec{n} \rangle = \delta_{ij}\sigma_i^2$. 
The data are then described by the total covariance matrix, $C= S + N$, which combines the covariance due 
to the signal $S$ and the covariance due to the noise, $N$.  The model also includes a set of linear parameters
to deal with the light curve means and long time scale trends (see the
Appendix).  In all cases we use the linear parameters to
remove the light curve mean. As we explore in \S\ref{sec:understanding}, long term trends can
affect the results for the LPVs. We experimented with removing polynomial trends up to third order, 
ultimately settling on using a cubic trend model for all stellar sources (LPVs, RR Lyrae, and Cepheids), and two trends for QSOs corresponding to the OGLE~III and IV portions of
the light curves because they can be offset.  We only used the OGLE
III light curves for the variable stars so there was no need to allow
for mean offset other than for the quasars. Our method automatically includes the uncertainties 
in the linear parameters into the uncertainties in the process parameters.

The final likelihood function in this framework is then 
\begin{equation} 
\label{eq:logl} 
\mathcal{L} = \frac{\exp\left(-(\vec{y} - L\hat{q})^TC^{-1}(\vec{y} - L\hat{q})/2\right)}{(2\pi)^{P/2}|C|^{1/2}|L^T C^{-1} L|^{1/2}}, 
\end{equation}
where $L\hat{q}$ is the best-fitting trend model described by the matrix, $L$, and the best-fitting linear coefficients, $\hat{q}$ 
(see the Appendix), $\vec{y}$ is the observed light curve, and $P$ is the number of parameters in the stochastic process model. 
The computational barrier to performing such an analysis lies in the
inversion of the covariance matrix, $C$. \cite{rybicki_press1992} 
found a method to reduce the scaling from $\mathcal{O}(N^3)$ to
$\mathcal{O}(N)$ for the cases $p = 1$ and $p = 2$, and this 
was the method used by \cite{kozlowski+2010} and \cite{macleod+2010} to rapidly analyse OGLE-III and SDSS light curves with 
the DRW model. \cite{ambikasaran+2014} has subsequently found an algorithm which is $\mathcal{O}(Np^2)$, for all $p$, and 
we use this sparse matrix method. The forecasting algorithms used by \cite{kelly_bechtold_siemiginowska2009, kelly+2014} 
are also $\mathcal{O}(N)$, but in \cite{kozlowski+2010} we found that evaluating the full likelihood given by Equation~\ref{eq:logl} 
had significantly more statistical power than the forecasting methods. 

We model the light curves by first finding best-fitting parameters that
maximize the likelihood (Equation~\ref{eq:logl}) using a simulated annealing
method, as implemented in
PyGMO\footnote{\url{https://github.com/esa/pagmo/tree/master/PyGMO}}. Our
problem necessitates a MLE approach for optimization as opposed to a
grid search because the length of the OGLE light curves requires large numbers of period
samples even before adding extra parameters. For a fixed phase error,
$\delta \phi$, the period has to be sampled by $\Delta P/P \sim
(\delta \phi/2\pi)(P/T)$, where $T \sim 3000$~days is the baseline of OGLE
light curves. For a short period Cepheid (RR Lyrae) 
period $P \sim 1$~day ($P \sim 0.1$~day) and $\delta \phi \sim 1$,
this requires $\Delta P/P \sim 1/3000$ $(1/30000)$. Using the MLE estimate for the best-fitting DRW or QPO model as a starting point, we then
sample the parameter space using the Python MCMC package \texttt{emcee} \citep{foreman-mackey+2013}. We use standard logarithmic 
priors for the process parameters. Our chain length is significantly larger than the autocorrelation time \citep[e.g.,][]{sokal1996} of 
the estimated parameters calculated by \texttt{emcee} for most light
curves. In this limit, the mean marginalized parameter estimated is
representative of the true mean of the parameter. Parameter errors are
quoted at 1-$\sigma$, and the best-fitting parameter is taken to be
the mean of the MCMC chains.  

To compare models with differing numbers of parameters, we use the Akaike Information Criterion (AIC) and the 
Bayesian Information Criterion (BIC). For a light curve with likelihood $\mathcal{L}$, $N$ data points, and $k$ 
model parameters, the Bayesian likelihoods (Equation~\ref{eq:logl}) are modified to be
\begin{equation} 
AIC \equiv -2\ln \mathcal{L} + 2k 
\end{equation}
\noindent and
\begin{equation} 
BIC \equiv -2\ln \mathcal{L} + k\ln N. 
\end{equation}
The BIC penalizes the addition of new parameters more heavily than does the AIC. These allow us to compare models with 
differing numbers of parameters, and, in particular, address the questions of which of the models (DRW or QPO) better 
fits the data, and whether a quasi-periodic model is a better fit than a periodic model. We present
the AIC results. Our initial criterion for rejecting sources better fit by white noise
(the limit $\tau \rightarrow 0$~days for either QPO or DRW) corresponds to rejecting objects with
 $\Delta AIC < -1$ between the white noise and the DRW/QPO models. In
order to compress the full range of the likelihood differences between
models $1$ and $2$, 
we show the distributions in terms of 
\begin{equation} 
f(x) = \hbox{sinh}^{-1} \left[\frac{AIC_{1} - AIC_{2}}{2}\right] = \hbox{sinh}^{-1}x,
\label{eq:f} 
\end{equation} 
This transformation provides a linear scaling for $|x| \la 1$ and transitions to a logarithmic 
scaling $\hbox{ln}|x|$ for $|x| >> 1$. The normalization is chosen such that the likelihood ratio 
is $95$ per cent at $f=\pm 1$ (corresponding to $\Delta \chi^2=4$ for
$\chi^2$ statistics). For $f < -1$ ($f > 1$), the source is best
described by model $1$ (model $2$) and at $|f| < 1$, the maximum 
likelihood probability for model $1$ (model $2$) is greater than $5$
per cent. The probability of the less favoured model plummets outside
the $|f| < 1$ region, falling below $0.07$ per cent for $|f| > 2$.

\section{Understanding the Results}
\label{sec:understanding}
We next discuss a series of experiments to help understand the global
results.  We start by simply confirming
that the process is internally consistent by generating QPO model
light curves with cadences, amplitudes
and uncertainties typical of the OGLE data and showing that we can
recover the input parameters
reasonably well.  We also show that we can recover the periods of the
LPVs with some dependence on
the treatment of long time scale trends in the light curves.
For shorter period variables, there is essentially never a problem in
estimating the period. When we then applied the method to the actual light curves, we found
that the
results showed a significant dependence on the light curve shape,
behaving as expected for relatively
sinusoidal light curves but with counter-intuitive results for very
non-sinusoidal variables like
long period Cepheids.  In particular, it was surprising to find that
QPO models of extremely
periodic RRab and long period Cepheids frequently preferred low
coherence QPO models or even
the DRW model.  This motivated a series of experiments modeling purely
periodic sine and
triangle waves as a function of the noise relative to the amplitude
that demonstrate the effect
of light curve shape on QPO parameter estimation.   Finally, we
introduce a
higher order, $p=4$, QPO model to confirm that a higher order model that can better mimic
the structure
of less sinusoidal light curves provides a means of addressing
the problem.

\begin{figure*}
\centering
\subfloat{
   \includegraphics[width=0.5\textwidth]{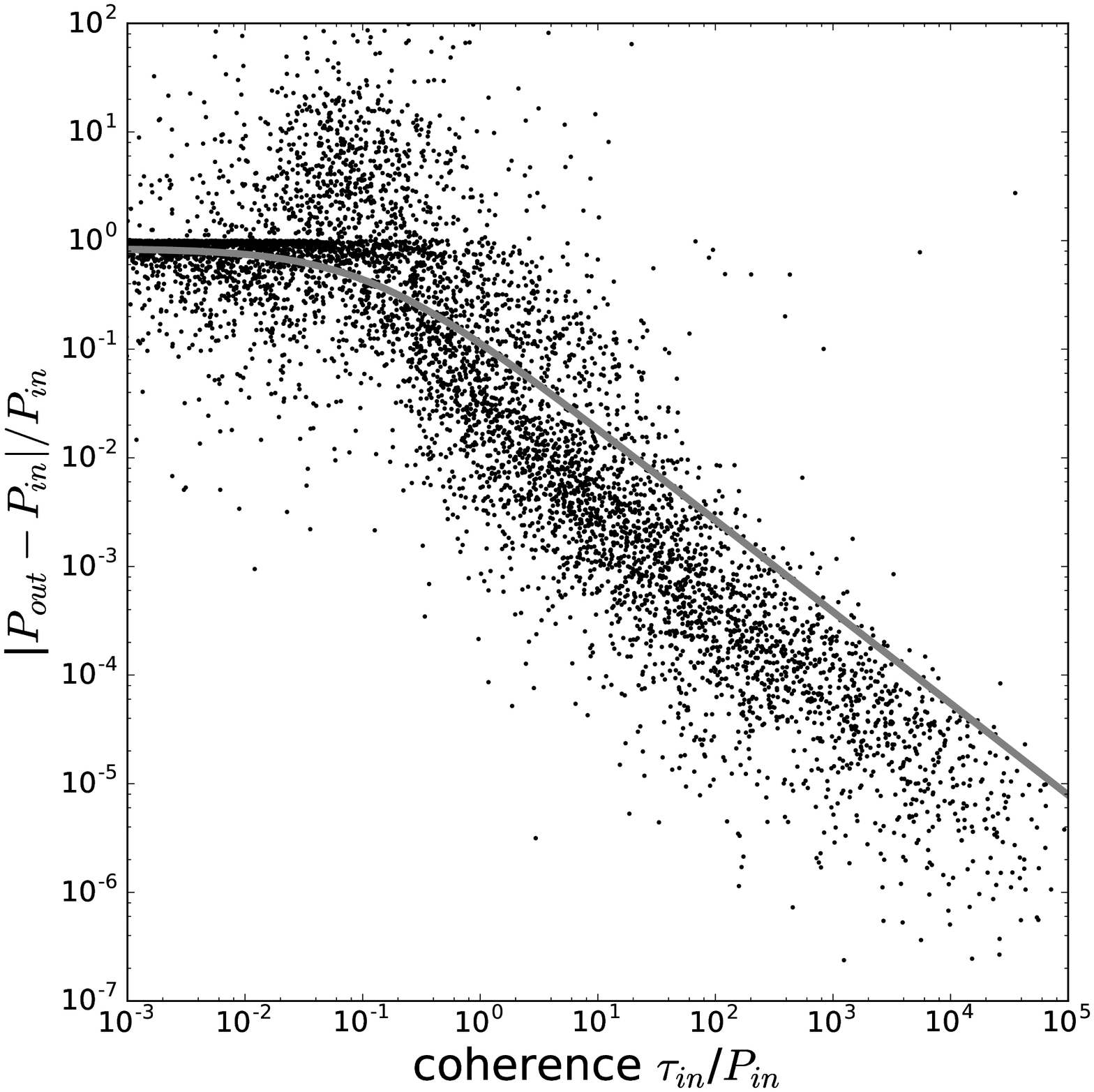}
}
\centering
\subfloat{\includegraphics[width=0.5\textwidth]{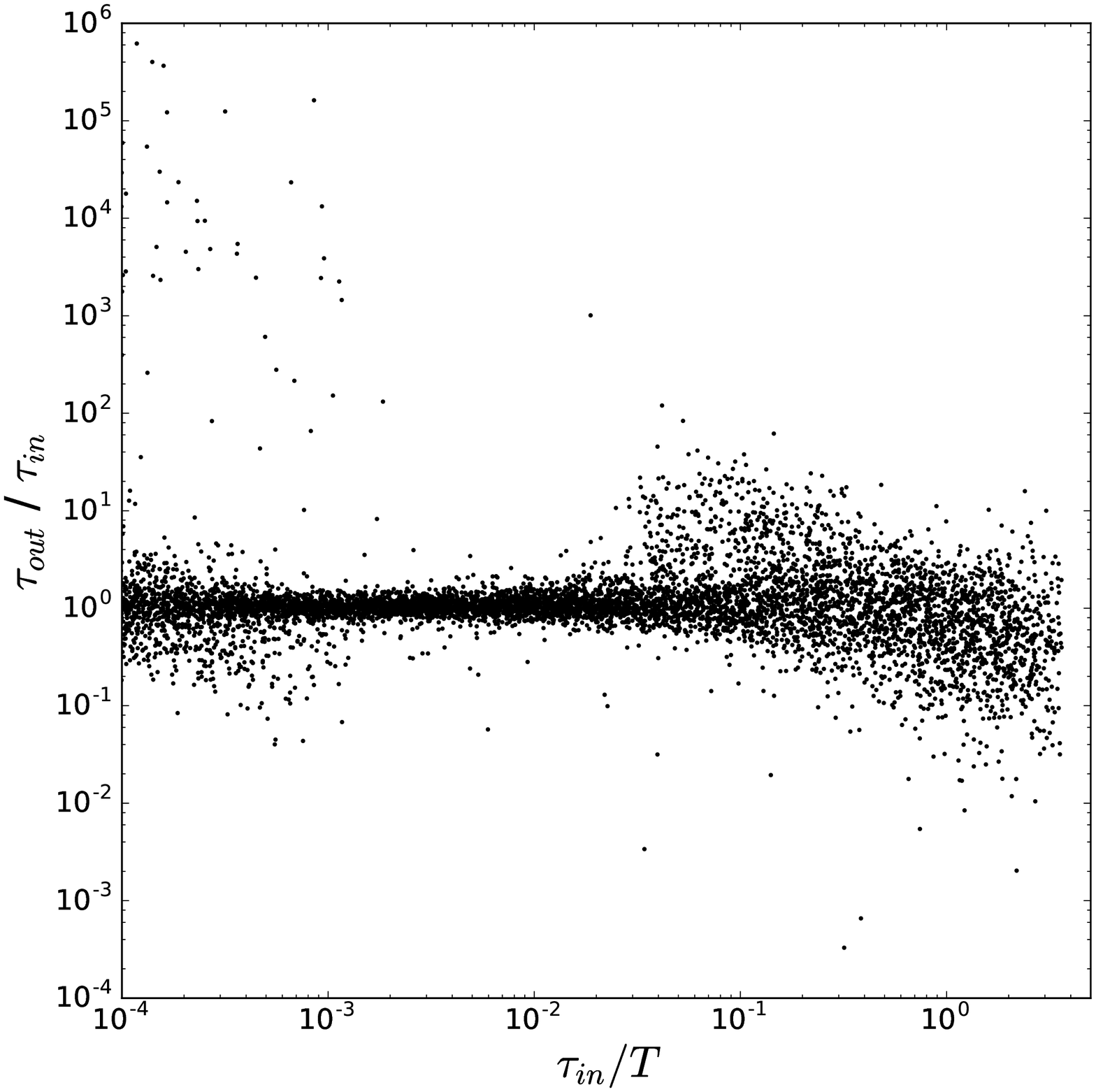}
}
\caption{Fractional error in period, $|P_{out} - P_{in}|/P_{in}$ as a function of the coherence of
  the oscillation, $\tau_{in}/P_{in}$, for artificial light curves (left).
  A simple model for the fractional error in period as a function
  of the coherence is shown as a solid line to guide the
  eye. Recovered versus input coherence
  time, $\tau_{out}/\tau_{in}$, as a function of the coherence time in
  units of the light curve baseline, $\tau_{in}/T$, where $T \approx
  3000$~days (right). Coherence times approaching the length of the
  light curve time baseline or shorter than the cadence ($\tau_{in}/T
  = 1\times10^{-3}$) are increasingly difficult to
  recover.}
\label{fig:Abaseline}
\end{figure*}
To provide a guide to interpreting the results in \S\ref{sec:results},
we first tested how well we can recover QPO parameters simply due to 
the modeling process itself.  We generated a sample of 12000
artificial QPO light curves using cadences typical 
of the OGLE light curves and with amplitudes, periods, and coherence time scales covering the range of the variables we 
study in \S\ref{sec:results}, with periods varying from $0.1$ to
$10^3$~days, $\tau$ varying from 0.1 to $10^6$~days, and amplitudes
varying from $0.01$ to $1$ mag. We used a fixed photometric uncertainty
of $0.02$~mag, which is roughly twice the 95$^{th}$ percentile of
the photometric error distribution of the OGLE light curves.  We generated the light curves using the Cholesky 
decomposition method outlined in \cite{zu+2013}.  If the covariance
matrix, $C$, can be Cholesky decomposed into 
$C = M M^T$, and $R$ is a vector of Gaussian random
deviates of unit amplitude, a light curve realization is simply $y = M
R$.

Figure~\ref{fig:Abaseline} illustrates the
results. Figure~\ref{fig:Abaseline}a shows that the fractional errors in the periods are well modeled by $| \Delta P|/P = 1/[B + (\tau/P)^{C}]$
with $B \simeq 14$ and $C \simeq 0.73$, as might be expected since the ``line width'' of the oscillation must
grow as the periodicity becomes less coherent. If the damping scale is
short compared to the overall light curve length, $\tau/T \lesssim
0.1$, then the median fractional uncertainty in $\tau$ is $\approx
10$ per cent, with a scatter of $\pm0.3$ dex (see Figure~\ref{fig:Abaseline}). The
uncertainty in $\tau$ becomes large as it approaches the light curve
timespan, $T$, generally in the sense that it can be greatly
over-estimated. That $\tau$ is
poorly recovered for $\tau/T \gtrsim 1$ 
justifies our definition of a purely periodic variable as one
where $\tau = 10^4$~days ($\tau
\approx 10T$) is the best model. We also demonstrate in
Figure~\ref{fig:Abaseline}b that $\tau$ is poorly
recovered for $\tau/T \lesssim 10^{-3}$, which is the typical sampling time-scale
($\sim 3$~days). Studies of DRW models for quasars
have also found that the damping time scale is difficult to recover
accurately from typically existing data where $\tau$ is
comparable to $T$ \citep{macleod+2010,kozlowski2016c}.

We also tested the ability of the QPO models to recover the published OGLE-III periods of the variable stars. 
We find that we can always recover the period of short-period
variables, with periods $\lesssim 100$~days. Care must be taken, however, to
adequately resolve period space when searching for periods much less
than the baseline of the light curve ($T \sim 3000$~days, for OGLE
light curves) because the likelihood peaks become increasingly narrow
due to the large number of periods spanned by the data.

\begin{figure*}
  \centerline{
     \includegraphics[width=0.5\textwidth]{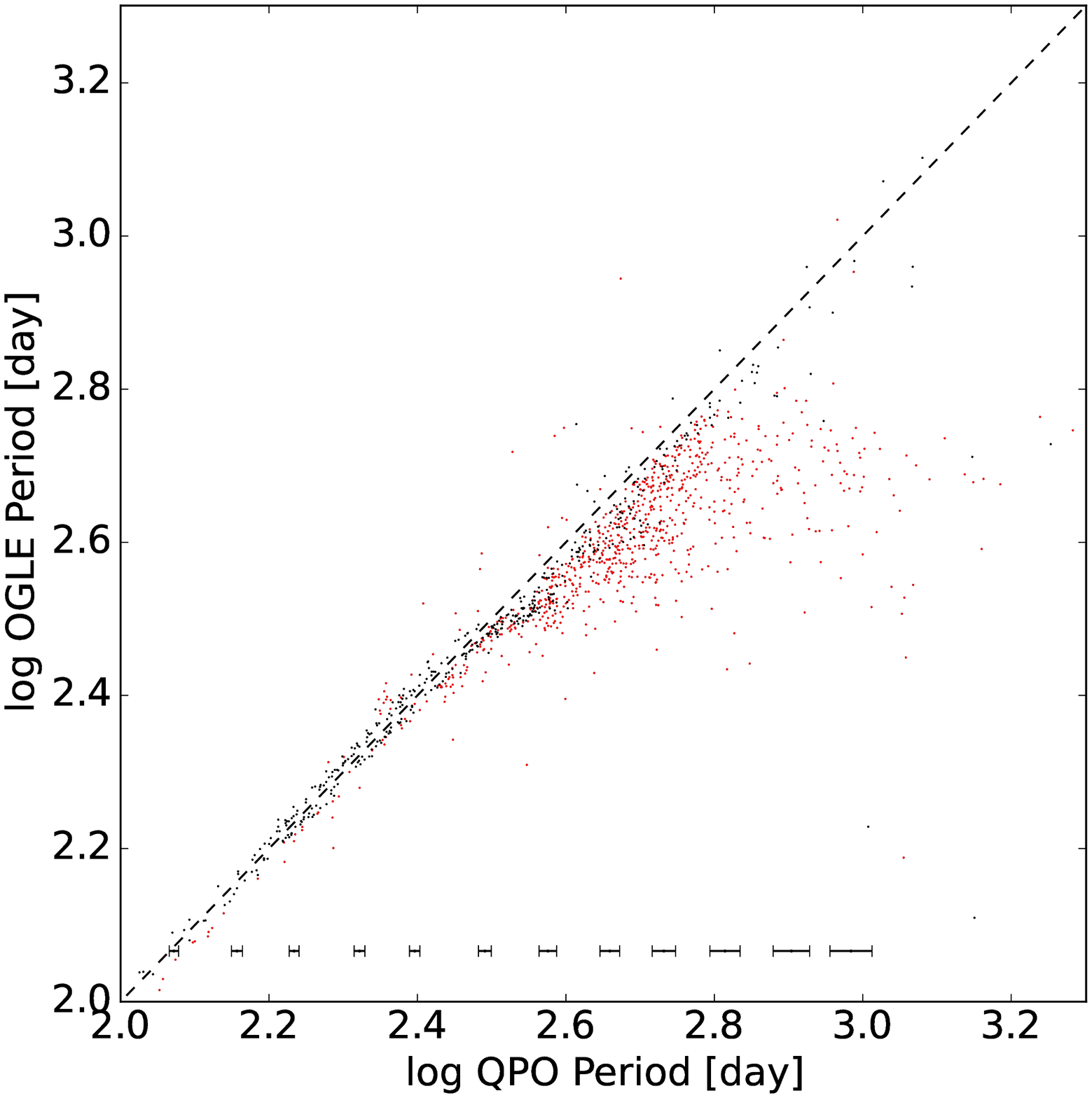}
     \includegraphics[width=0.5\textwidth]{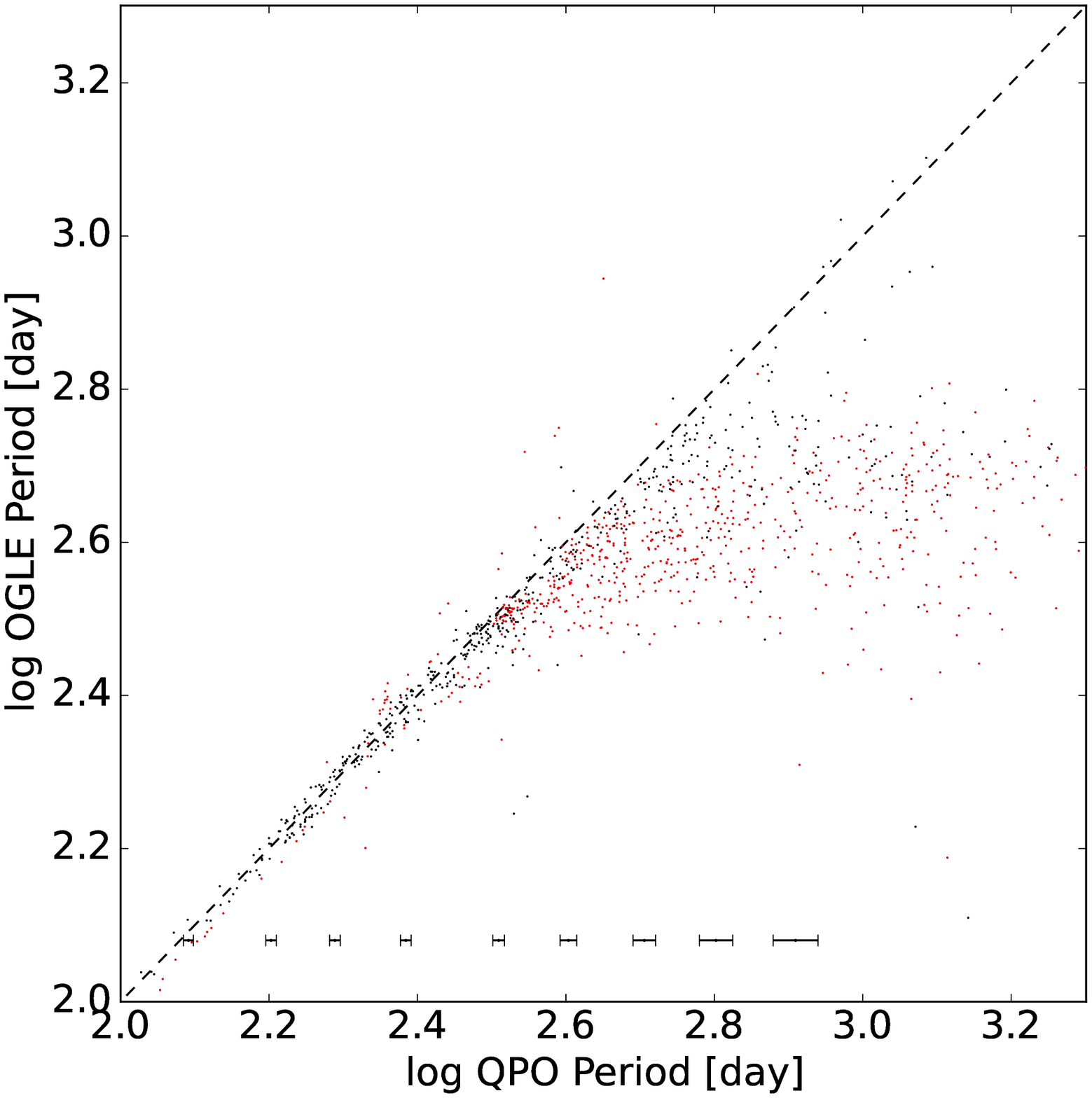}
     }
\caption{A comparison of QPO and OGLE period estimates for Miras using
  either cubic (left) or constant (right) temporal trend models.
  The diagonal line represents a one-to-one correspondence and the error bars along the bottom indicate the median QPO
  period uncertainties for a series of logarithmic bins in period.  The red points in each panel are the $3\sigma$
  period outliers in the other panel.  In general the cubic trend model provides better agreement and is adopted
  as our standard model.
  }
\label{fig:p_v_p}
\end{figure*}

\begin{figure*}
  \centerline{
     \includegraphics[width=0.5\textwidth]{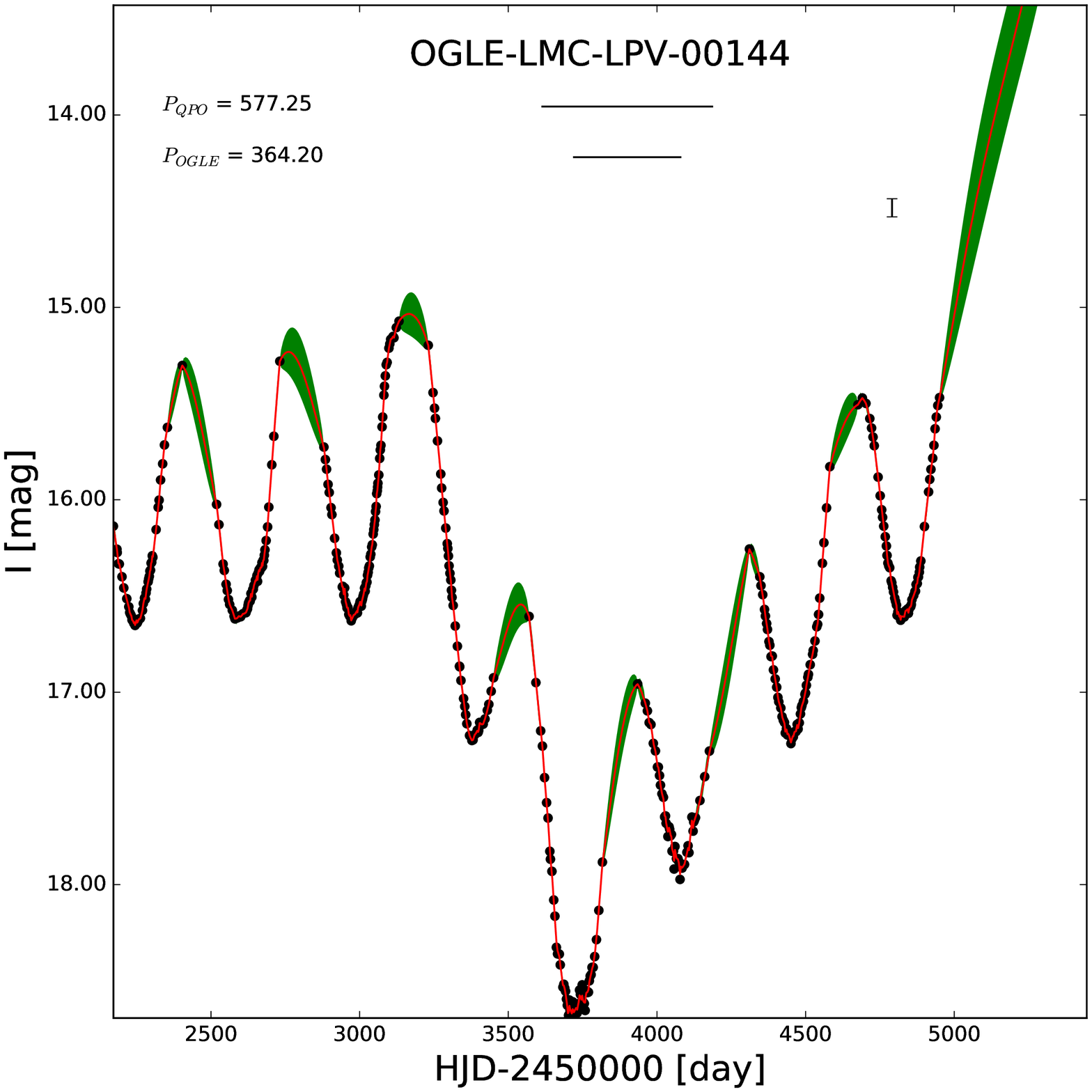}
     \includegraphics[width=0.5\textwidth]{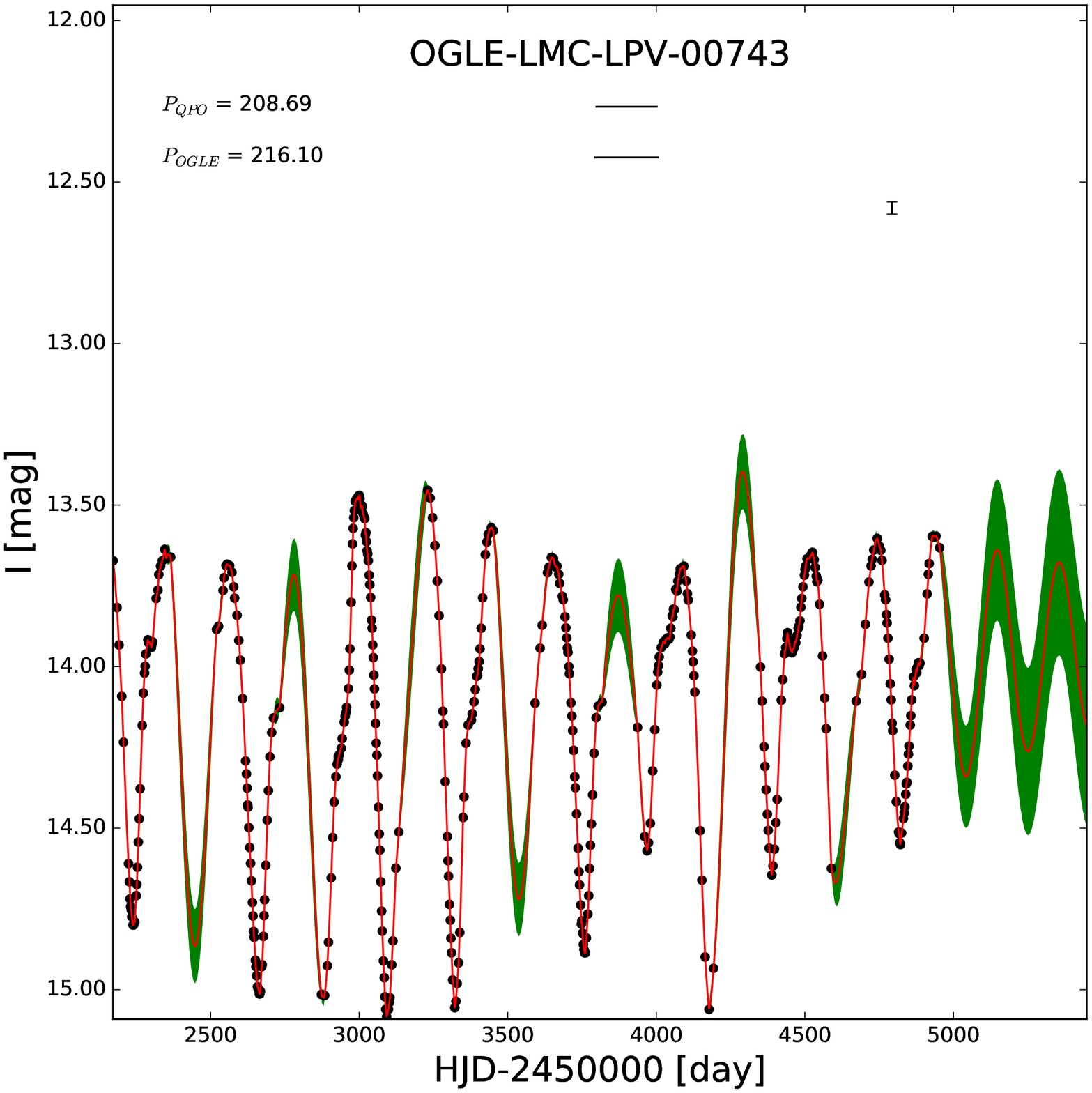}
     }
\caption{Miras with large (left) and small (right) period differences
  between the OGLE and the QPO models.  The
   points are the OGLE data and the curve is the average of QPO models at the indicated, fixed $P_{QPO}$ that
   fit the data well.  The green ``error snake'' represents the $1\sigma$ dispersion of these light curve models
   around the mean.  Median photometric errors are shown as a single vertical bar.
   The periods are listed in each panel along with lines to graphically show the scale of
   the periods and their differences relative to the light curves. The
   divergence of the light curve model in the left panel is a
   consequence of the cubic trend model.
   }
\label{fig:miras_p}
\end{figure*}

As shown in Figure~\ref{fig:p_v_p} for the Miras, the situation is quite different for the LPVs.  
The period agreement is very 
good for shorter periods ($P < 300$~days), but there are significant
numbers of cases where the QPO period estimate is 
significantly longer than the OGLE period.  It is not surprising that the discrepancies are concentrated 
at longer periods, since even for truly periodic sources the fractional period uncertainty will be on 
the order of the period divided by the time span of the data.  However, the discrepancies are larger 
than we would expect based on our simulations of modeling randomly generated QPO light curves (see Figure~\ref{fig:Abaseline}b).  
Figure~\ref{fig:miras_p} shows the light curves and QPO models for two  
examples, LPV-00144 where the period discrepancy is large ($P_{OGLE}=364$ versus $P_{QPO}=577$~days) and  
LPV-00743 where the period discrepancy is smaller ($P_{OGLE}=216$ versus $P_{QPO}=209$~days). 
The obvious 
distinction is that LPV-00743 lacks the large amplitude changes of
LPV-00144. Such large scale features are common among the period
outliers. 

This suggested that the period differences 
might in part be driven by the cubic model for long term trends that is included in our standard analysis, 
so we repeated the calculations removing constant (light curve mean), linear and quadratic trends.  Figure~\ref{fig:p_v_p}  
compares the results for simply removing a constant to the default
cubic model, and we see that the number of outliers increases 
significantly.  We also show in Figure~\ref{fig:p_v_p} where outliers for the two trend models lie in  
the distribution found for the other trend model. Given the results of
these experiments, we kept the cubic model as our default for stellar
sources. The trend model choice is also important for the recovery of
DRW parameters, particularly for QSOs. Unsurprisingly, fitting trend models of higher order
than a simple mean results in removing genuine stochastic behavior and
biases the estimates of $\tau$ toward artificially small or high
values (not shown). For this reason, we opted to only remove a mean
each for the OGLE-III and OGLE-IV sections of the QSO light curves for the DRW model. A cubic was removed for each of the two light curve sections when fitting QPO models to QSOs. 
 
One technical point to emphasize about Figure~\ref{fig:miras_p} is that the QPO model curve is not a particular 
``best-fitting'' light curve, but the statistical average of light curves consistent with the data (see the 
discussion in \citealt{rybicki_press1992}).  The ``error snake'' surrounding this curve is the $1\sigma$ dispersion 
of these light curves around their mean.  A particular statistical realization of a light curve consistent 
with the data would track the mean with deviations statistically bounded by the ``error snake'', and the 
construction of such constrained light curve realizations is also described in \cite{rybicki_press1992}.  These 
constrained realizations will show more small scale structure than the mean light curve, and this is a 
large part of the variance captured by the ``error snake''.

As we developed our analysis, we found (see \S\ref{sec:results}) that
periodic variables with light curves that strongly deviated from
sinusoids (e.g., fundamental
mode Cepheids and ab-type RR Lyrae) were frequently either favouring
QPO models with short coherence times or even the DRW model over the
coherent, periodic limit of the QPO model. Figure~\ref{fig:bestfit} 
illustrates the issue for the RRab variable OGLE-LMC-RRLYR-00001,
which is clearly a coherent periodic variable over the OGLE baseline,
but has a best-fitting QPO
coherence time of $\tau \sim 30$~days for a period of $P \simeq 0.64$~days
so that $\tau/P \sim 40$ instead of $\tau/P \sim \infty$.

\begin{figure*}
\centering
\subfloat{
   \centering
   \includegraphics[width=0.5\textwidth]{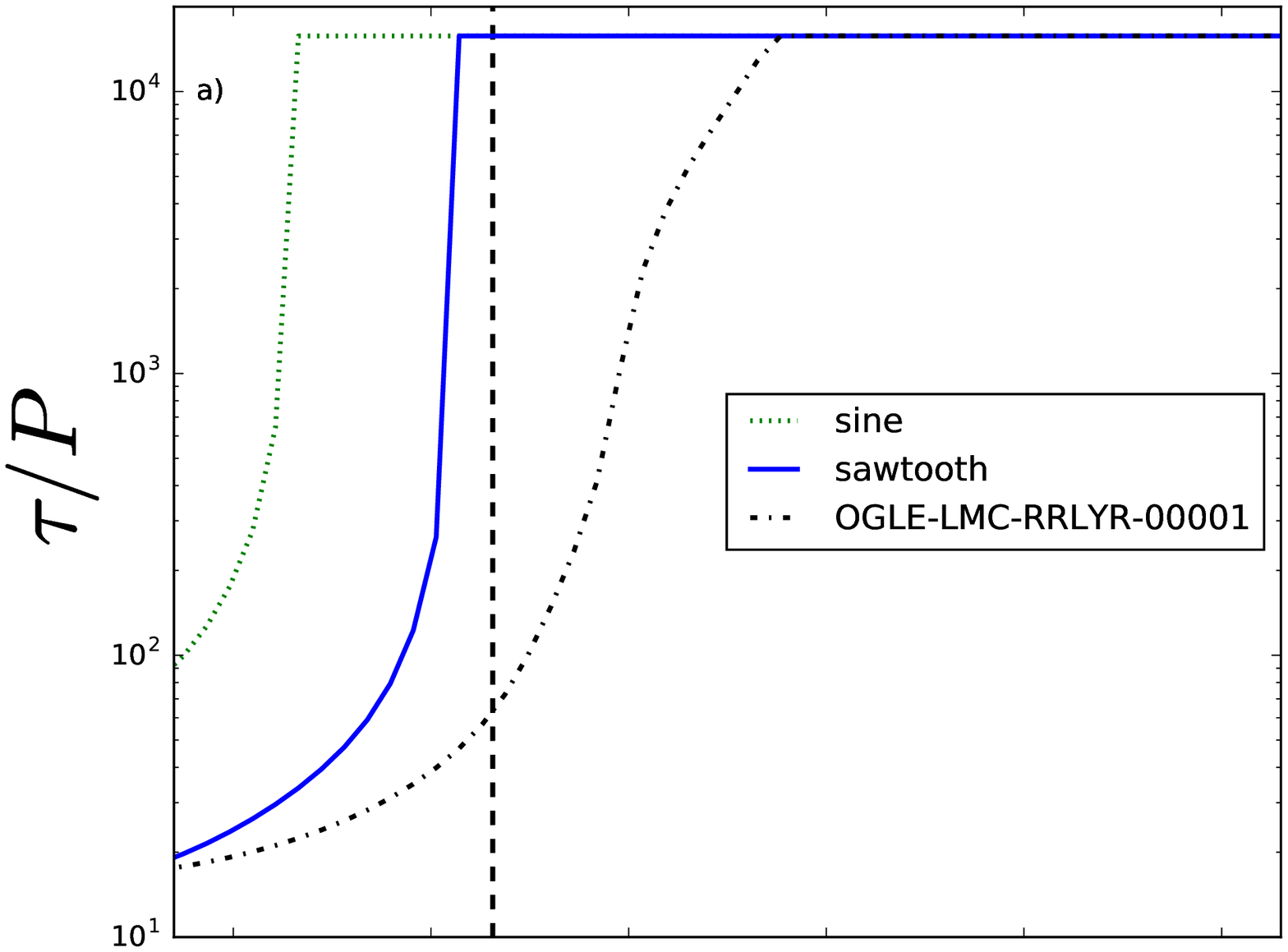}
}
\subfloat{
  \centering
  \includegraphics[width=0.5\textwidth]{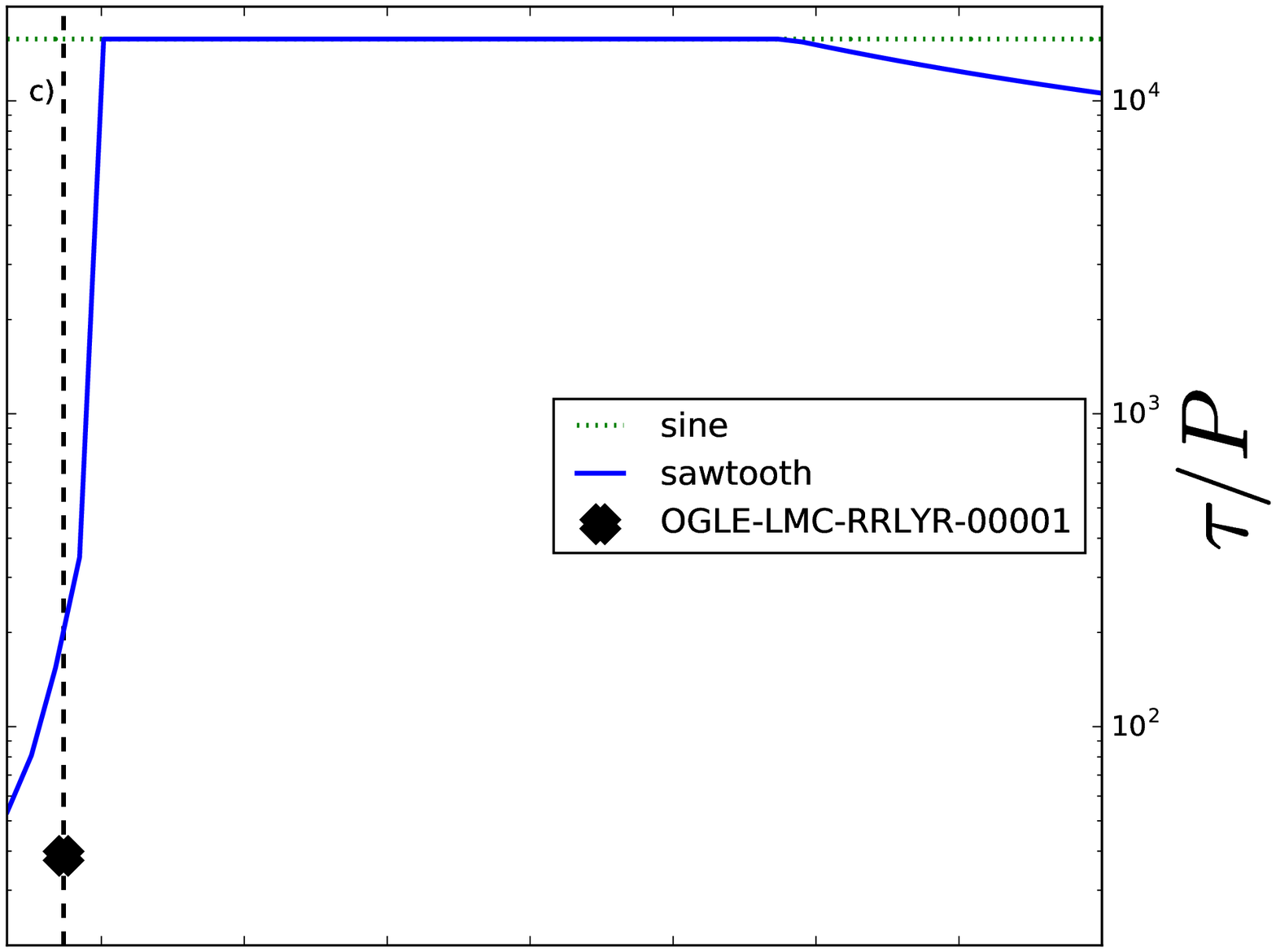}
}\\[-3.2ex]
\subfloat{
   \centering
   \includegraphics[width=0.5\textwidth]{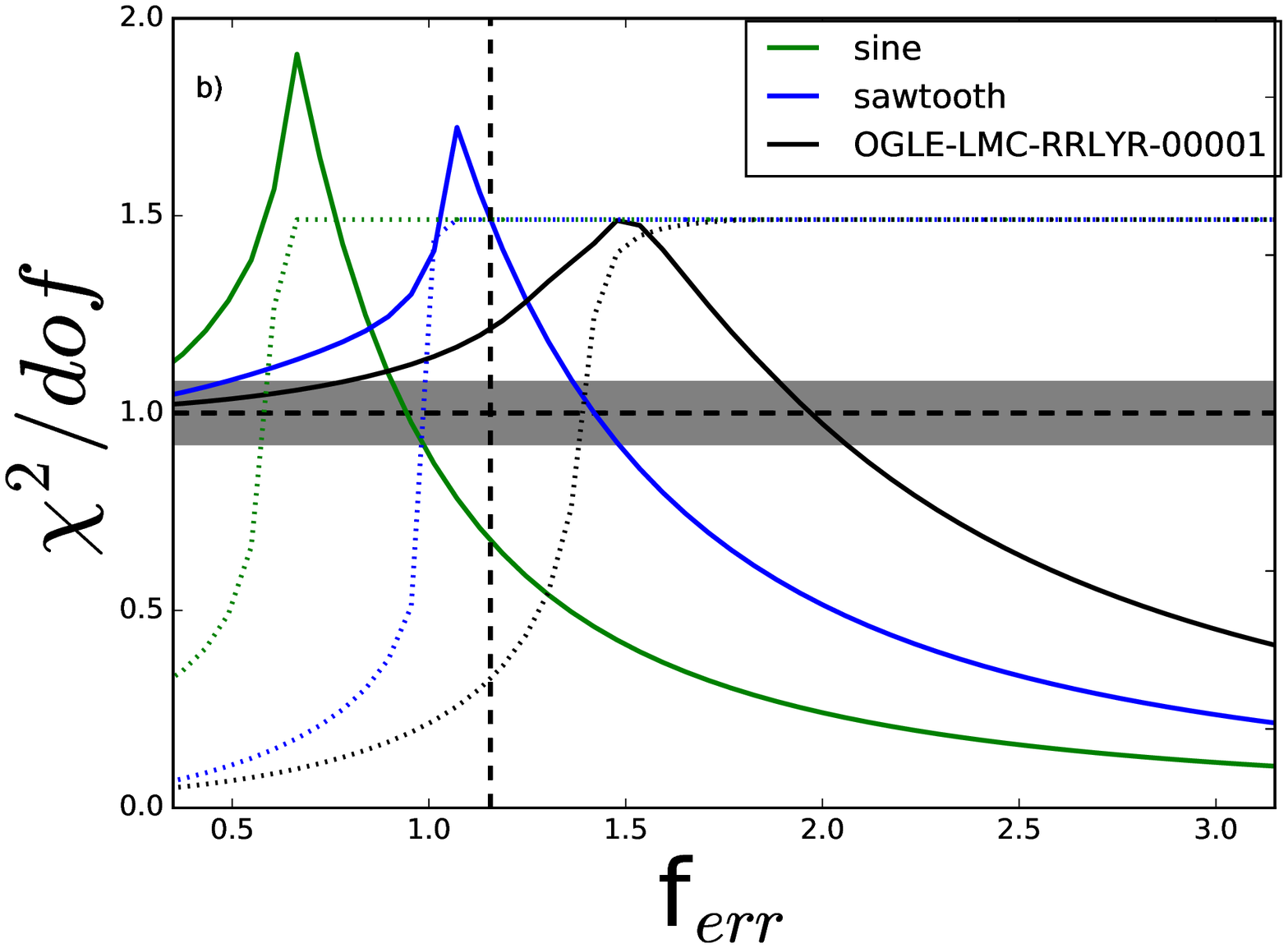}
}
\subfloat{
  \centering
  \includegraphics[width=0.5\textwidth]{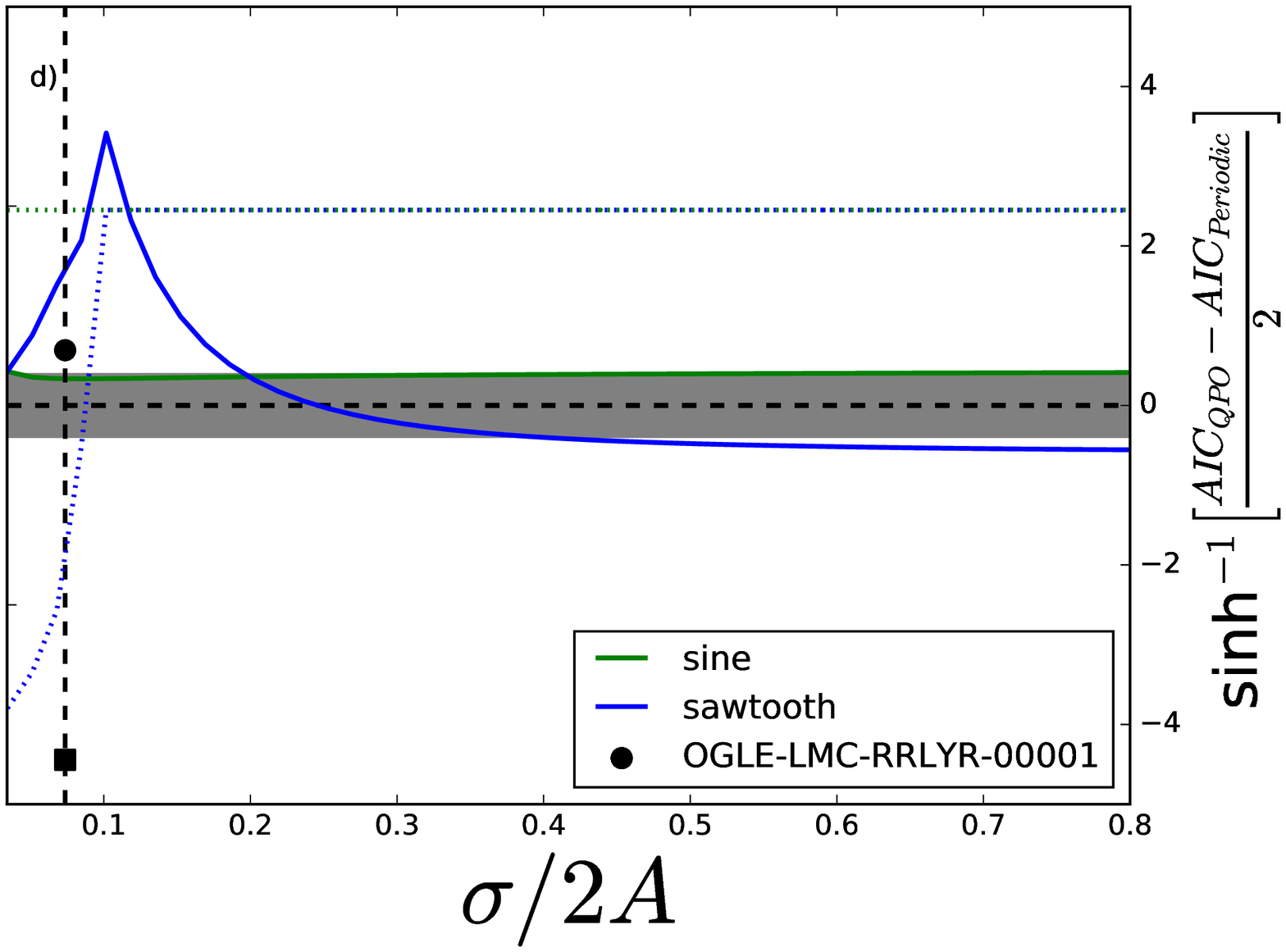}
}
\caption{The upper panels show the coherence ratio, $\tau/P$, for the OGLE RR Lyrae
  OGLE-LMC-RRLYR-00001 and artificial light curves with sinusoidal or
  sawtooth waveforms. The lower panels show the QPO model preference
  (Equation~\ref{eq:f}; dotted lines, right scale) and the
  $\chi^2/dof$ (solid lines, left scale); positive/negative values
  favour periodic/QPO. The vertical dashed line corresponds to the
  recommended factor by which the reported OGLE-III errors for
  OGLE-LMC-RRLYR-00001 should be increased
  \protect\citep{wyrzykowski+2009}. The horizontal dashed line
  corresponds to a $\chi^2/dof$ of unity and the grey band indicates
  the expected $68$ per cent confidence region assuming Gaussian
  statistics. The left columns show these quantities as a function of
  the error scaling factor $f_{err}$ (error bars $\sigma \rightarrow
  f_{err} \sigma$). The right column shows these quantities as a
  function of the photometric error relative to the peak-to-peak
  amplitude, $\sigma_i/2A$. A point marks the position of the
  $\chi^2/dof$ for OGLE-LMC-RRLYR-00001, and a square marks its model preference.}
\label{fig:sawtooth}
\end{figure*}

To test the hypothesis that light curve shape could bias coherence
time estimates and model preference, we modeled purely periodic sinusoid and
triangle wave variability. Artificial light curves
were generated using the same I-band magnitude ($18.8$~mag), period, cadence, and photometric error
$\sigma_i=0.1$~mag as the variable OGLE-LMC-RRLYR-00001 shown in Figure~\ref{fig:bestfit}. We want to consider two possible effects. The
first is the consequence of misunderstanding the photometric uncertainties and the
second is the consequence of the ``signal-to-noise'' ratio of the
variability. Figure~\ref{fig:sawtooth}a shows the effect on the
recovered $\tau$ of
rescaling the reported noise by a factor of $f_{err}$, meaning that we analyse
light curves with the error defined to be $f_{err}\sigma_i$ instead of
the true error, $\sigma_i$. The light curve itself is not
modified. Under-estimating the true error should drive the solution
towards $\tau \rightarrow 0$~days since this is the limit corresponding to
photometric white noise, while over-estimating the error should do
the reverse. As we see in Figure~\ref{fig:sawtooth}a, the actual
variable and the two model variables follow this basic trend. For the
sinusoid, the best solution is periodic even when the uncertainties
are deliberately under-estimated, while the sawtooth requires slightly
over-estimated errors to be seen as a periodic oscillator. The actual RR Lyra approaches the periodic
solution if the true uncertainties are $f_{err} = 1.5$--$2.0$ times the
raw OGLE estimates rather than the recommended scaling by $f_{err}
\approx 1.2$ \citep{wyrzykowski+2009}.

Figure~\ref{fig:sawtooth}b shows model preference
(right axis) and the $\chi^2/dof$ (left axis) as a function of
$f_{err}$. Model preference is determined by the AIC criterion and the
scaling defined in Equation~\ref{eq:f}, for which a positive (negative) value indicates
a preference for the periodic (QPO) model. The sine, sawtooth,
and OGLE-LMC-RRLYR-00001 light curves all converge to a periodic model
for a sufficiently large $f_{err}$, essentially tracking the $\tau$
parameter estimates. The goodness of fit shows two branches. If the
errors are greatly under-estimated ($f_{err} \rightarrow 0$), the data
can be well-modeled by the white noise limit of the QPO process ($\tau
\rightarrow 0$~days). If the errors are over-estimated, the data can be
over-fit ($\chi^2/dof < 1$) and the process can be periodic ($\tau
\rightarrow \infty$~days). For the sinusoid, we see that for $f_{err} = 1$, we have
$\chi^2/dof = 1$ and $\tau \rightarrow 10^4$~days, as we would expect. For
the sawtooth, the fit at $f_{err} = 1$ is poor ($\chi^2/dof \simeq
1.4$), the best-fitting model is only moderately coherent ($\tau
\approx 170$~days), and the QPO model is favoured even though the light
curve is periodic. If we increase $f_{err}$ to $f_{err} \approx 1.4$,
we can make $\chi^2/dof \simeq 1$, $\tau \rightarrow 10^4$~days, and
have a
preference for the periodic model. The RR Lyra behaves
similarly to the sawtooth, but would need $f_{err} \simeq 2$ to
achieve $\chi^2/dof = 1$. The non-sinusoidal light curve shape
requires the assumption of additional white noise (higher values of $f_{err}$)
for the preferred solution to be periodic.

In the second experiment, we generated light curves with different ratios
$\sigma/2A$ between the errors and the peak-to-peak variability
amplitude, $2A$. Figures~\ref{fig:sawtooth}c
and~\ref{fig:sawtooth}d show the best-fitting coherence ratio,
$\tau/P$, the goodness of fit $\chi^2/dof$, and the degree of
preference for the QPO or periodic models for this case. The sinusoidal model
behaves exactly as expected, always finding the limit $\tau
\rightarrow 10^4$~days, with $\chi^2/dof \simeq 1$, and an
overwhelming preference for the periodic solution. The sawtooth
model, on the other hand, prefers low coherence QPO models at high
signal-to-noise and then switches to preferring the periodic solution
at low signal-to-noise. In the transition between these two limits at
intermediate signal-to-noise, the fits to the data are poor, with
$\chi^2/dof$ significantly greater than unity. OGLE-LMC-RRLYR-00001, whose parameters are indicated with
an `x', behaves like a sawtooth light curve.

We can explain these behaviors by thinking about the problem in terms
of the power spectrum. The QPO model produces a Lorentzian power
spectrum $P(\omega) = \sigma^2 / \sqrt{8\pi} \sum_\pm \left[ \tau^{-2} + (\omega\pm
  \omega_0)^2 \right]^{-1}$  for a period of $P= 2\pi/\omega_0$ and a
damping time scale $\tau$.  A sinusoidal light curve becomes a pair of
delta functions at $\pm \omega_0$.   A white noise spectrum is simply
a constant independent of $\omega$.   The spectrum of a non-sinusoidal
periodic variable is a series of delta functions at frequencies $\pm n
\omega_0$, $n=1,2,...$, with the envelope of amplitudes representing the Fourier
series representation of the light curve shape.

When we fit a periodic triangle wave in the periodic limit of the QPO,
we can match only the first, $n=1$ peak at $|\omega|=\omega_0$ in the
power spectrum, leaving all the power in the $n>1$ peaks.  We can
obtain a better fit by increasing the error estimates, which
corresponds to adding additional white noise. This works best when the
signal-to-noise ratio of the triangle wave is low because white noise
is not a very good representation of the $n>1$ peaks.  For the noise
fixed to the correct level, the power in the extra peaks can also be
partially captured by giving the model a finite $\tau$.  Suppose, for
example, that the ratio of the power in the $n=1$ and $n=2$ peaks is
$R_2$, then the QPO model can have the same power at the second peak
if $\tau \sim P R_2^{-1/2}$.   So the more the light curve shape
deviates from truly sinusoidal, the more the best-fitting model will
prefer a finite $\tau$ even though the underlying model is actually
periodic.

This strongly suggests testing a higher order QPO model with the
second period fixed to be $P/2$ (i.e. the $n=2$ term). Periodic
variables, such as Cepheids and RR Lyrae, with strongly non-sinusoidal
light curves, can be well fit by Fourier series with
period-dependent amplitudes \citep[see,
  e.g.,][]{pejcha_kochanek2012}, and therefore should be better fit by
a QPO model with a $n=2$ term. A simple second order
example of such a light curve, $f(t) = A\cos (2\pi t/P) + B \cos (4
\pi t/P +\phi),$ corresponds to the periodic limit of a higher order
($p=4$) stochastic process with the
periods fixed in a 2:1 ratio. In \S\ref{sec:results} we will compare QPO results for the single component
($p=2$) model to the higher-order $p=4$ QPO model results
for RR Lyrae and Cepheids with the primary period fixed to the primary
OGLE period and the secondary Fourier period fixed to half this period, fitting only an amplitude of oscillation for the
primary and secondary Fourier period, and a shared coherence
time ($A$, $B$, and $\tau$). The phase $\phi$ does not affect the
autocorrelation function.
 
\section{Results} 
\label{sec:results} 
Keeping these issues in mind, we now examine the results from fitting
all the variables with the DRW and QPO models. We first explore
whether the QPO model is a better description of periodic and
quasi-periodic phenomena than the DRW model. Next we examine how the
QPO model distinguishes between periodic and quasi-periodic
phenomena. Finally, we examine the QPO parameter distributions for the
various variable classes with an eye towards classification.

\begin{figure*} 
\centering
\includegraphics[width=\textwidth]{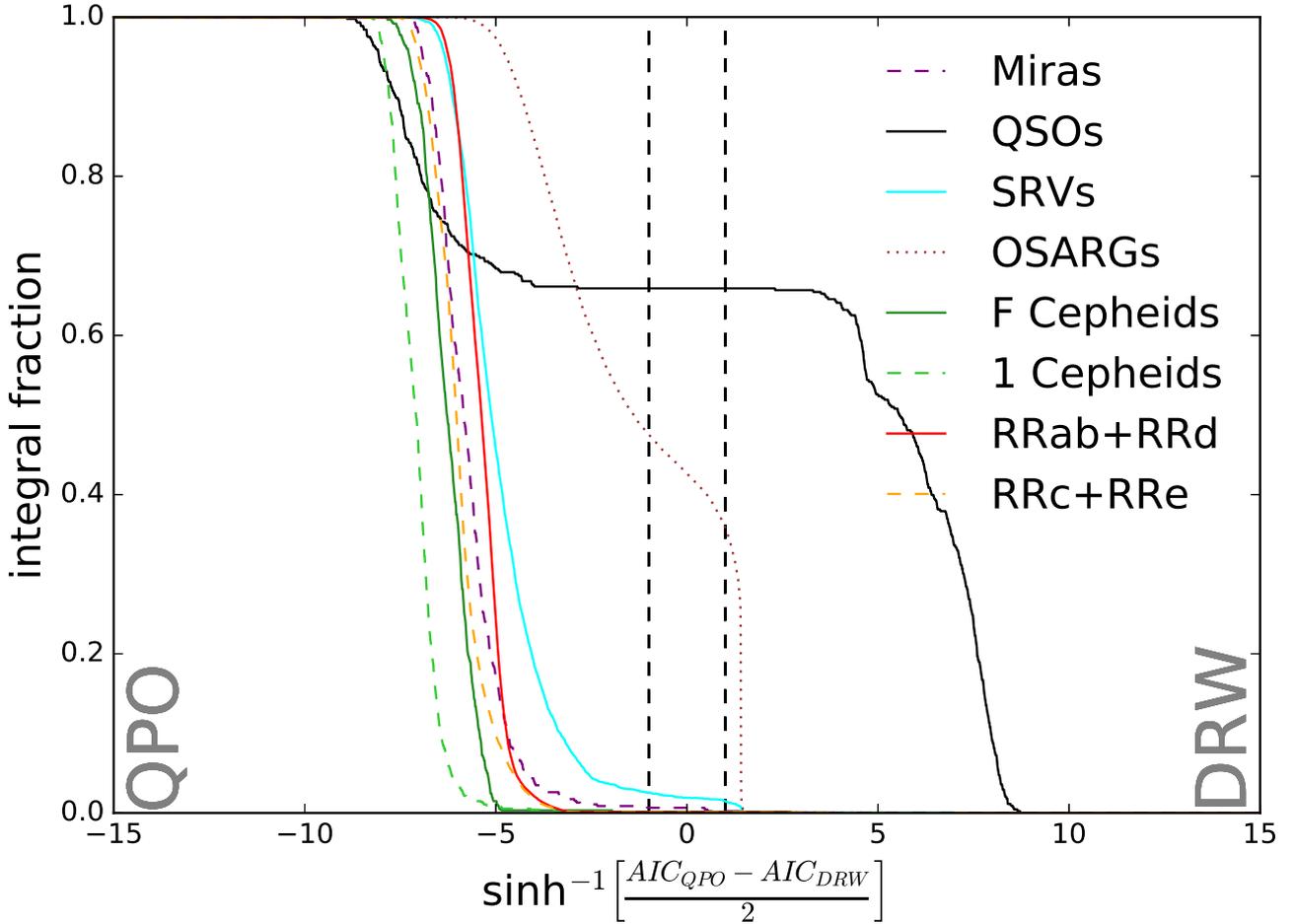}
\caption{The integral distribution of the labeled variable classes in
  $\sinh^{-1}\left[(\text{AIC}_{QPO} - \text{AIC}_{DRW})/2\right]$,
  where the QPO (DRW) model is favoured for negative (positive) values.
  The $\sinh^{-1}x$ mapping  is linear for $x \ll 1$ and then
  logarithmic for large $x$, so that the full dynamic range of the
  likelihood ratio can be displayed.  The mapping is normalized so
  that the two vertical lines at $|x|=1$ correspond to 95 per cent confidence likelihood ratios in favour of the QPO ($x=-1$) or DRW ($x=1$) models. The BIC distributions are very similar.}
\label{fig:aicbic}
\end{figure*}

\subsection{Discriminating between the QPO and DRW models} 
\label{sec:qpo_v_drw}
Figure~\ref{fig:aicbic} shows  
the distribution of the variable classes in the relative probabilities
of the QPO and DRW models for the AIC likelihoods. We use the
$\sinh ^{-1}x$ scaling of Equation~\ref{eq:f} to compress the
full dynamic range into the figure -- when $|x| = 1$ the associated
model is favoured at $95$ per cent confidence.
 
\begin{figure*} 
\centering 
\centerline{ 
   \includegraphics[width=0.4\textwidth]{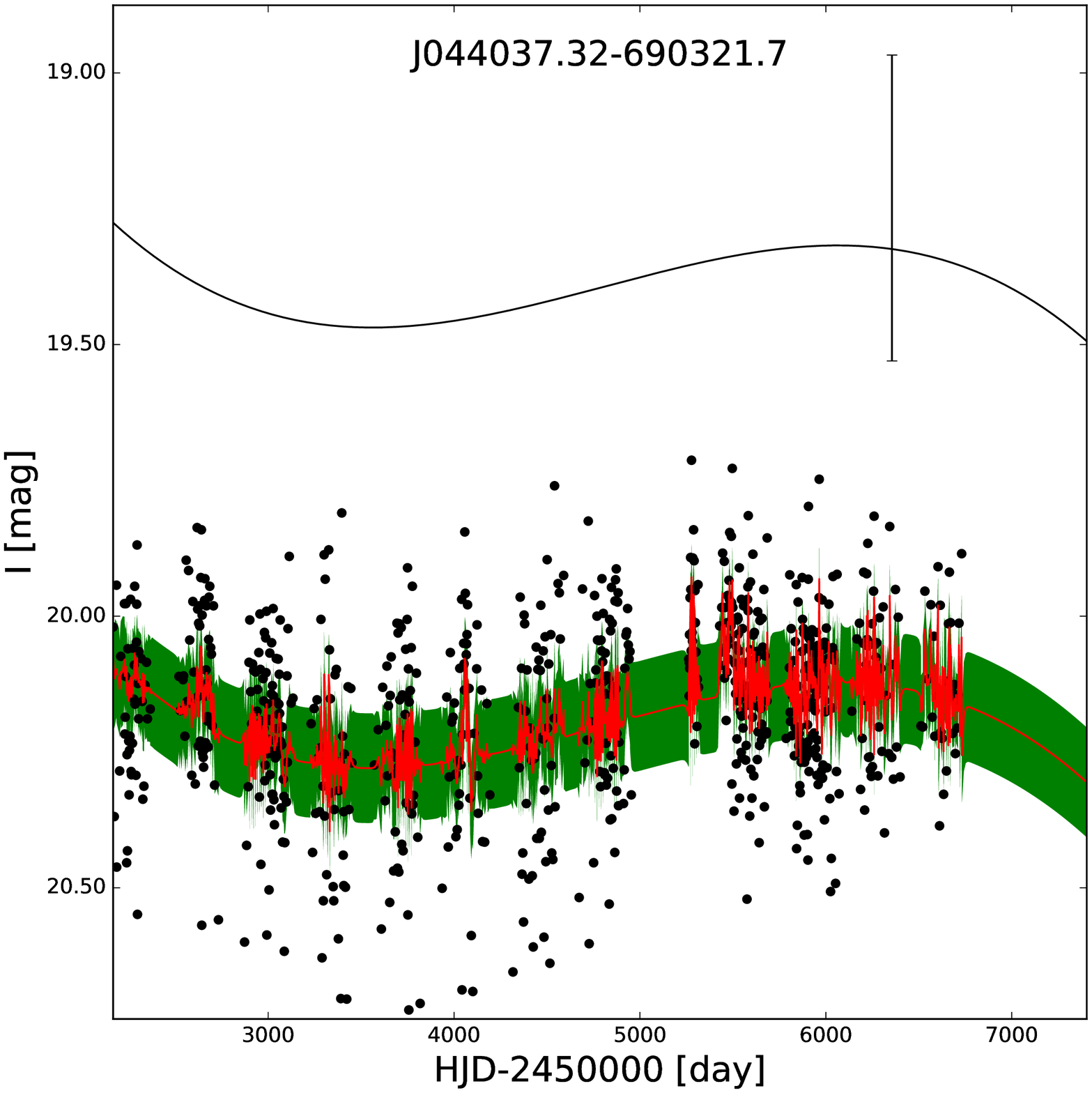}
   \includegraphics[width=0.4\textwidth]{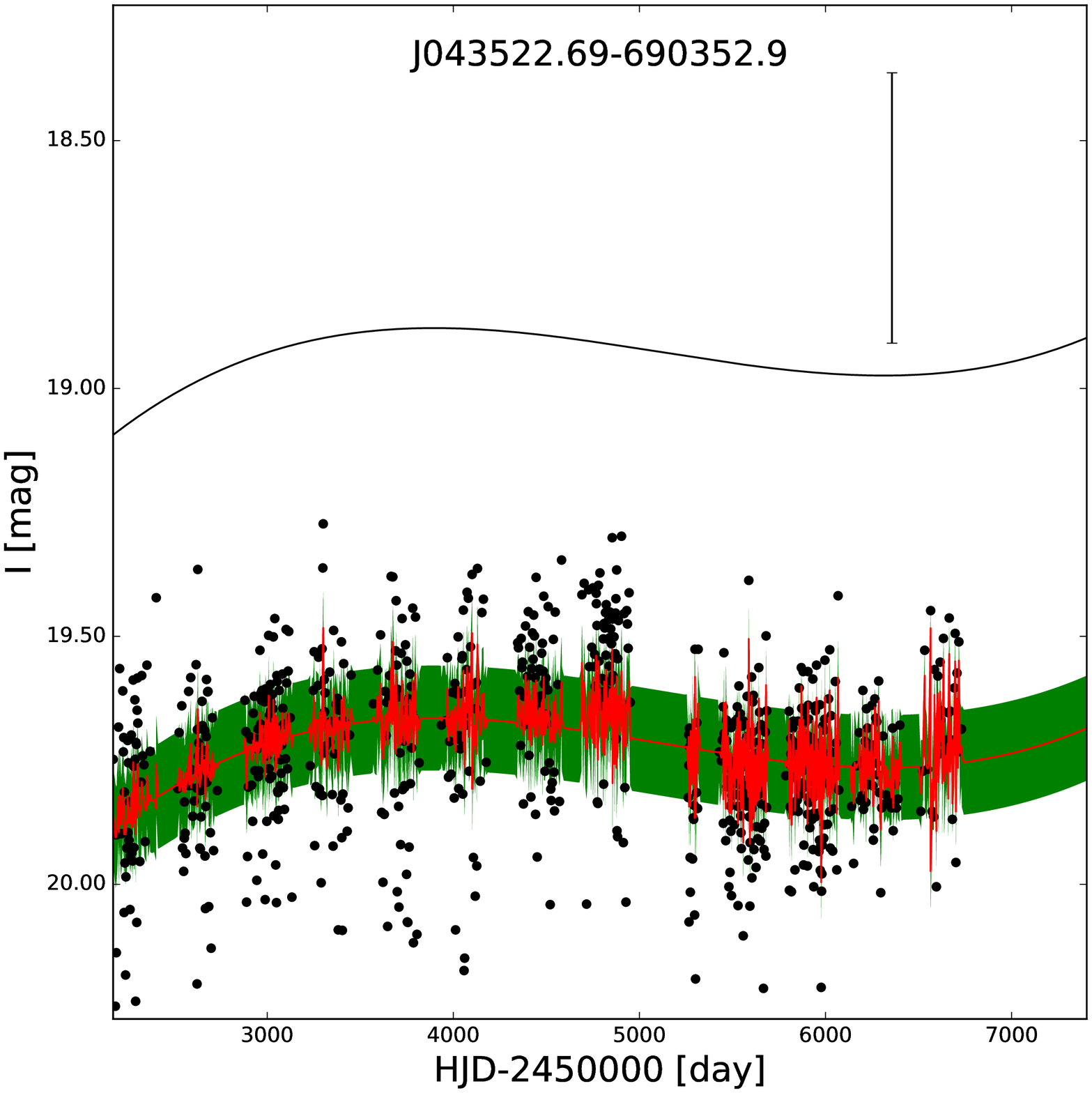}
   } 
\centerline{ 
   \includegraphics[width=0.4\textwidth]{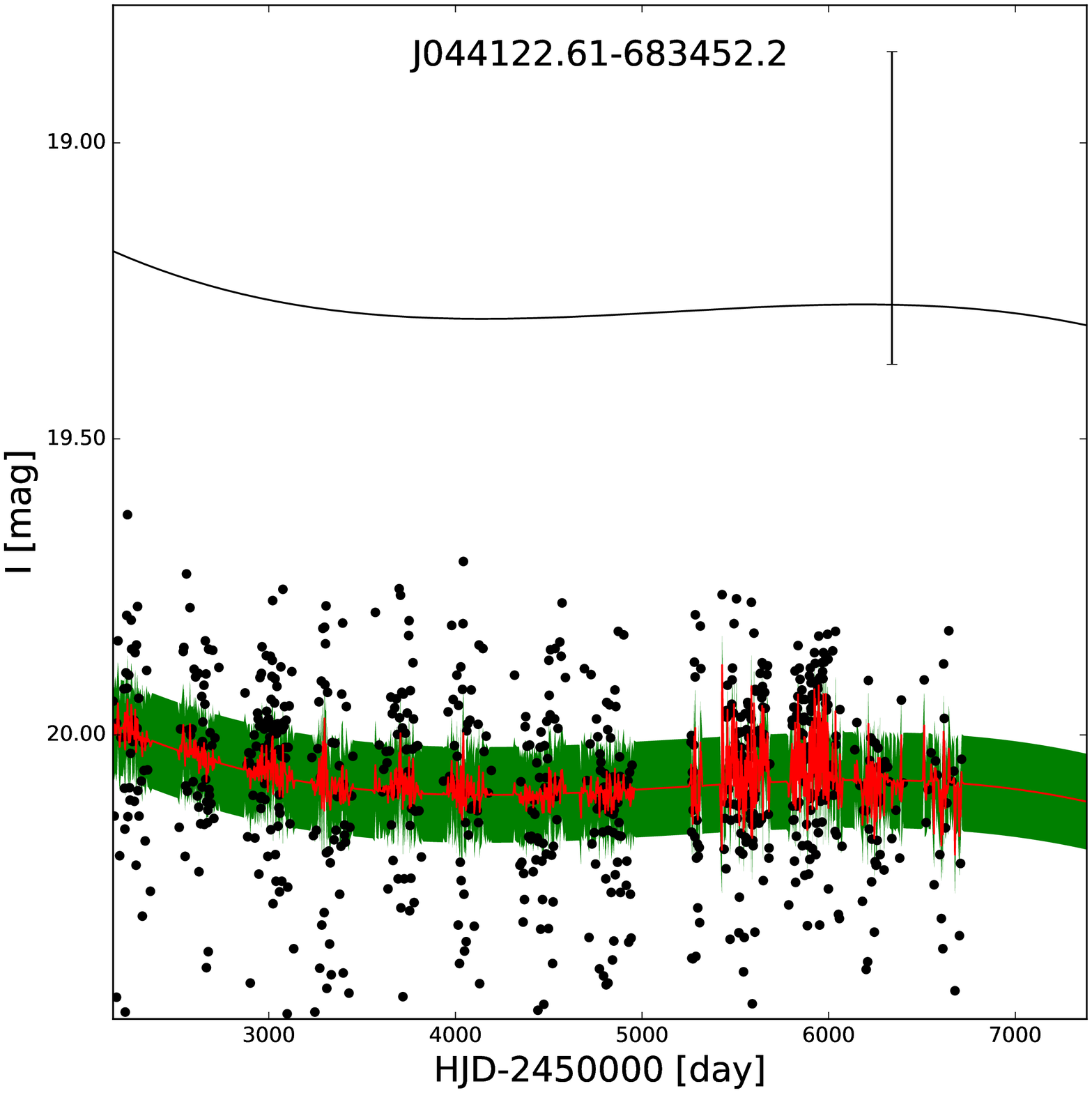}
   \includegraphics[width=0.4\textwidth]{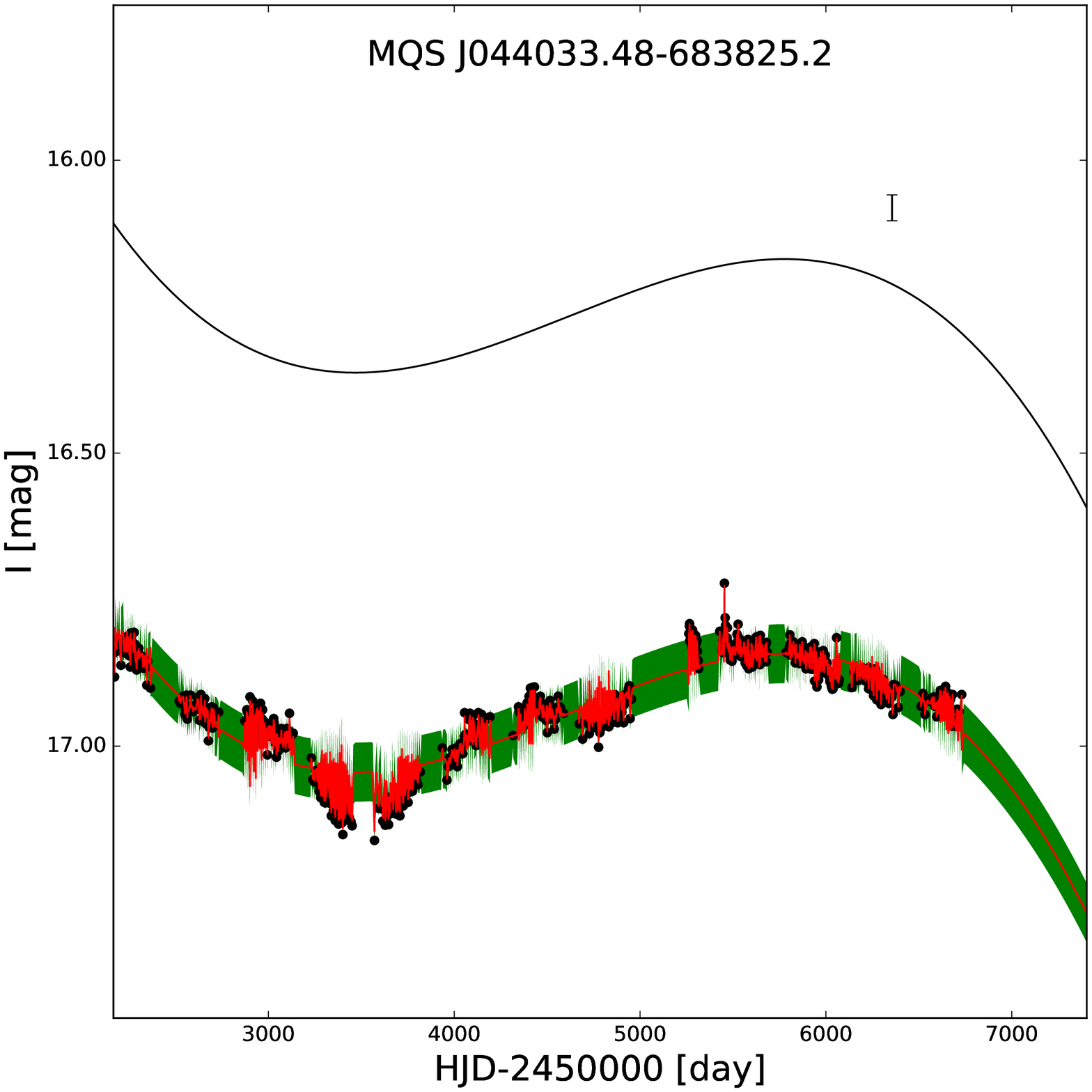}
   } 
\caption{Four examples of QSOs for which the QPO model was favoured over the DRW model.  The points 
  are the OGLE data and the error bar indicates the median photometric
  error. The offset curves show the cubic trend models. The curve and shaded regions again correspond to the average of the best-fitting
  light curves and their dispersion (see
  \S\protect\ref{sec:understanding}). All these examples have QPO periods smaller than the
  OGLE observing cadence, and are nominally coherent, but with a coherence time that is also
  smaller than the OGLE cadence. Essentially, the cubic trend model
  has captured the true variability and the QPO model is acting as a
  contribution to the noise.
  } 
\label{fig:noise_qso}
\end{figure*}

Quasar variability is generally believed to be stochastic rather than quasi-periodic, and we find that a 
 majority of the quasars (58 per cent) strongly favour the DRW model
 ($x>1$), leaving 42 per cent of the QSOs that 
prefer the QPO model.
This is broadly consistent with the earlier study of \cite{andrae_kim_bailer-jones2013}, who found that the DRW best 
described 80 per cent of the 6304 QSOs in their sample, while 27 per cent preferred a simultaneous DRW and sinusoid model. 
While the authors did not consider the CARMA(2,1) QPO model, their
finding that a significant fraction were consistent with a purely 
periodic sinusoid plus a DRW seems consistent with our results. 
Figure~\ref{fig:noise_qso} shows four examples of QSO light curves that strongly favour the QPO 
model.  They are all quasars where the light curves are dominated by long term trends over the
time span of the data, with 22 per cent also having very poor signal-to-noise ratios.  One problem with the MQS quasar sample is that  
many of the quasars are relatively faint for the OGLE survey, and so lack high quality  
light curves.

In reality, the preference of most quasars for the QPO model is illusory because the
best-fitting QPO model is also ``incoherent'' in the sense that $\tau/P
<1$. Essentially a cubic trend
model plus a long period, relatively incoherent oscillation can fit
slow quasar variability well in such cases. If we add the requirement that the QPO model
must not only be preferred by the information criterion, but must also have a best-fitting QPO model with $\tau/P>1$, then the fraction of ``quasi-periodic'' quasars drops
below 20 per cent. Even among these objects, however, half have a coherence
time less than the OGLE cadence. In this regime,
the QPO model differs little from the white-noise limit of a DRW
model: the period is meaningless because the coherence time is so short
that its effect is to broaden the photometric errors. These solutions are found in spite of
evident variability on time scales greater than the cadence. Essentially the variability is largely modeled by the cubic
trend, and the QPO is acting as an additional source of
white noise (see, e.g., Figure~\ref{fig:noise_qso}).

As one would expect, most of the variable stars in Figure~\ref{fig:aicbic} are  
better fit by the QPO model. RR Lyrae and Cepheids overwhelmingly show
a strong preference ($f < -1$) for the QPO over the DRW model. The
LPVs prefer a QPO over the DRW in similar proportions. A
small fraction of the RR Lyrae ($0.02$ per cent) and Cepheids ($0.1$
per cent)
have a strong preference ($f > 1$) for DRW. Only $0.4$ per cent of RR Lyrae
and $0.3$ per cent of Cepheids are incoherent ($\tau/P < 1$). The results
with a higher-order QPO model introduced in \S\ref{sec:understanding}
are similar, but the higher-order models eliminate the ``tail'' of
systems towards a preference for the DRW model, particularly for RRab
and RRd sources whose light curves are not well described by
sinusoids. 

The largest population of DRW-preferring variable stars is the OSARGs. Almost
a quarter of the OSARGs in our sample prefer the DRW model. The best-fitting QPO models for the OSARGs better modeled by the
DRW model almost always have $\tau/P < 1$. In the limit of incoherent
oscillations, the QPO model increasingly resembles the DRW model, and
the fits become indistinguishable by their likelihoods and
$\chi^2/dof$. In this limit, the AIC model preference will
always favour a DRW model, because it has fewer parameters. Although periods may remain
detectable in a periodogram for marginally incoherent sources, the DRW model can still be statistically preferred
in the AIC formalism due to its fewer degrees of freedom. An issue
to keep in mind for the OSARGs is that their variability amplitudes are comparable to their photometric
noise, which makes it difficult to distinguish among models.
 
\begin{figure*} 
\centering
  \subfloat{
    \includegraphics[width=0.5\textwidth]{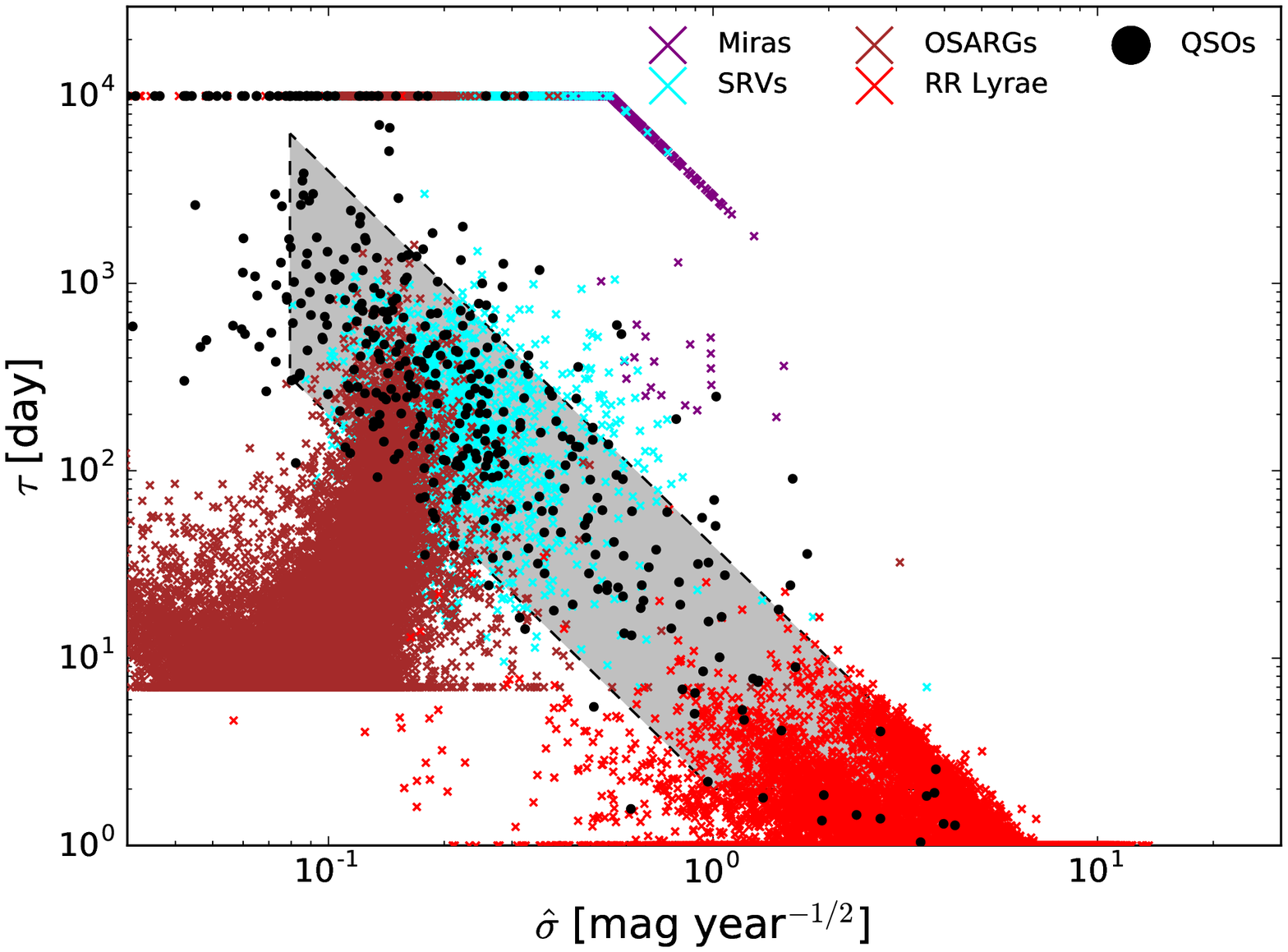}
}
\centering
\subfloat{
    \includegraphics[width=0.5\textwidth]{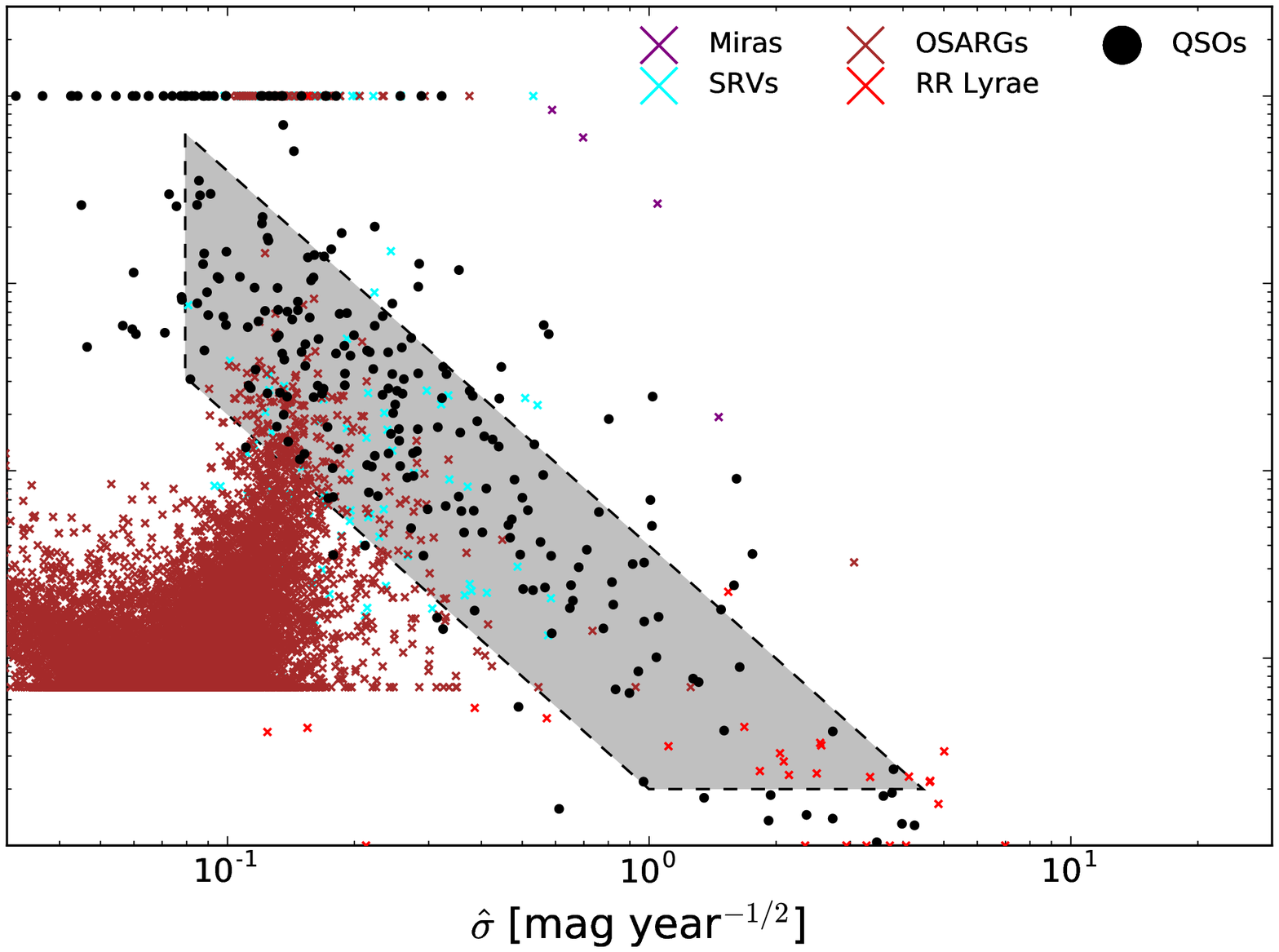}
     }
\caption{Distribution of several variable classes and QSOs in the DRW parameter space of
  $\hat{\sigma} \equiv \sigma_{DRW}\sqrt{2/\tau}$ and $\tau$ to match
  the equivalent figures in \protect\cite{kozlowski+2010}. The left panel
  shows all objects, while the right panel shows the distribution
  after eliminating sources that prefer ($x<0$ in
  Figure~\protect\ref{fig:aicbic}) the QPO model. The grey shaded region is
  the quasar selection region proposed by
  \protect\cite{kozlowski+2010}. Given the uncertainties in estimating
  $\tau$, the optimal selection region for QSOs
  will vary based on the time baseline of a typical light curve. For
  our purposes, we retain the selection region based
  on OGLE-III time baselines. Imposing the cut on the relative likelihoods
  of the DRW and QPO models eliminates $\sim 7$ per cent of the quasars but
  $67$ per cent of the variables.}
\label{fig:drw} 
\end{figure*} 

Figure~\ref{fig:drw} reprises the similar figure in \cite{kozlowski+2010}, comparing the 
distribution of quasars and variable stars in DRW parameters.  If no other criteria are imposed (left panel), the 
selection region proposed by \cite{kozlowski+2010} contains 62 per cent of the quasars, but also contains 0.52 
variable stars for every quasar.

In practice, \cite{kozlowski+2010} considered additional simple
selection criteria (e.g., color and magnitude) that significantly
increase the efficiency of the selection process, but here we only
consider the time domain criteria. If we impose the criterion that the light curves must also prefer 
the DRW model, so $AIC_{QPO} - AIC_{DRW} > 0$, the contamination is greatly reduced (the right panel of 
Figure~\ref{fig:drw}).  The selection region now contains 55 per cent of the quasars, but only 
0.20 variable stars for every quasar.  The additional criterion loses
only 7 per cent of the quasars but  
reduces the stellar contamination by 67 per cent. We attempted to increase
the efficiency of quasar selection by also selecting objects that had
incoherent QPO models. However, this does not improve quasar 
selection because the largest contaminant are OSARGs, which also favour
incoherent QPO models. The MQS quasars are, of course, selected 
from a region of very high stellar density (the Magellanic Clouds!), which makes the density of variable 
stars tremendously larger than in a typical extragalactic field ($\sim
1$ QSO per $10^5$ stars in the Magellanic Clouds!).  In a true
extragalactic field, the  
purity of such a variability-selected sample would be far
higher. Additional criteria such as the magnitude and color criteria
considered by \cite{kozlowski+2010} could also be developed to avoid the
issues leading to quasars favouring the QPO model. For example, we
could also require more variability ``power'' in the stochastic process
than in the cubic polynomial or require a minimum variability power.
 
\begin{figure*} 
\centerline{ 
  \includegraphics[width=\textwidth]{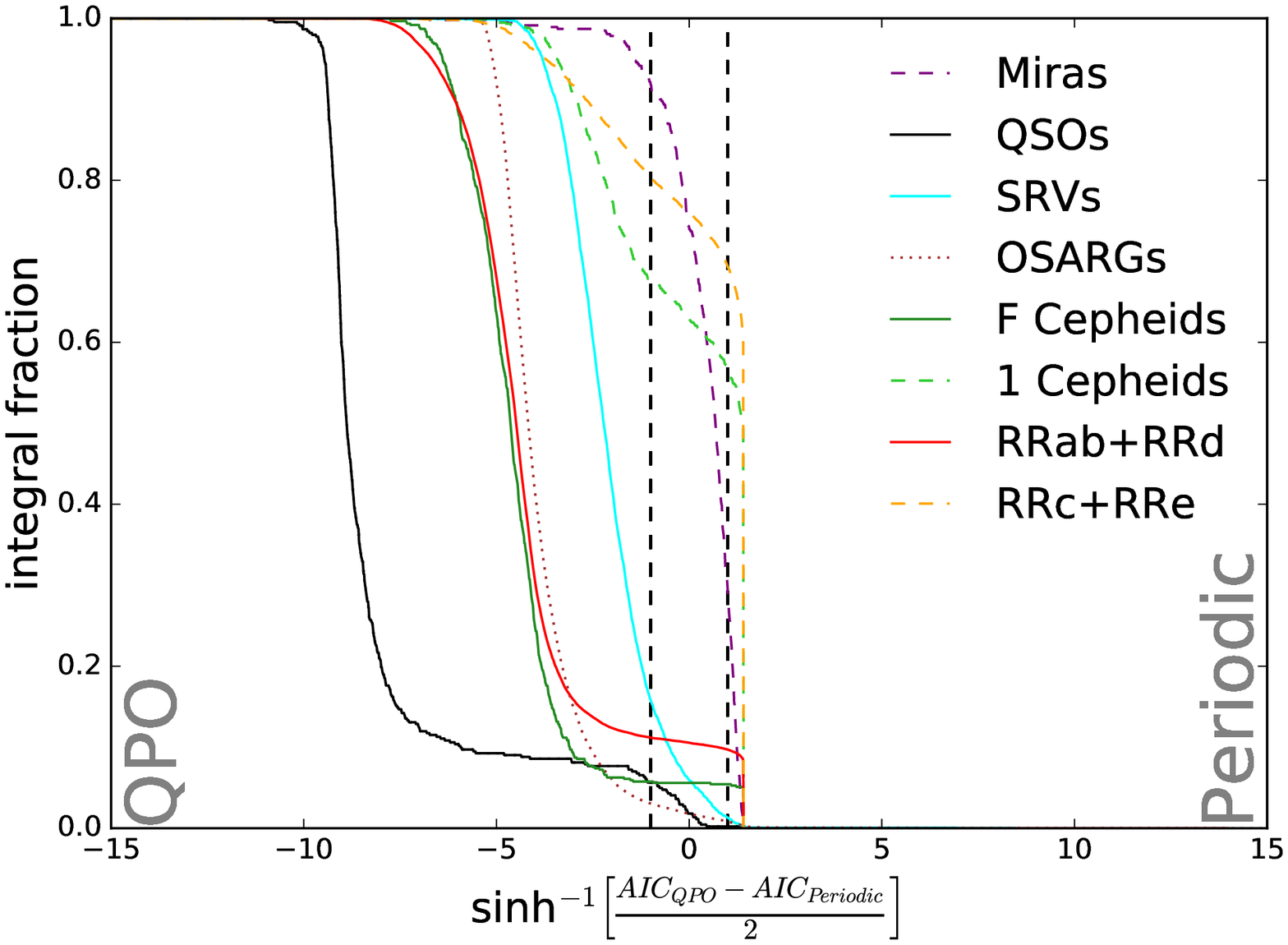}
  } 
\caption{Distribution of selected variables in the AIC likelihood ratio $\sinh^{-1}x$ with 
    $x=(AIC_{QPO}-AIC_P)/2$ for favouring periodic ($x>0$, right) or
  quasi-periodic ($x<0$, left) variability.   
       } 
\label{fig:p_v_quasi_p} 
\end{figure*} 

\begin{figure*} 
\centerline{ 
  \includegraphics[width=\textwidth]{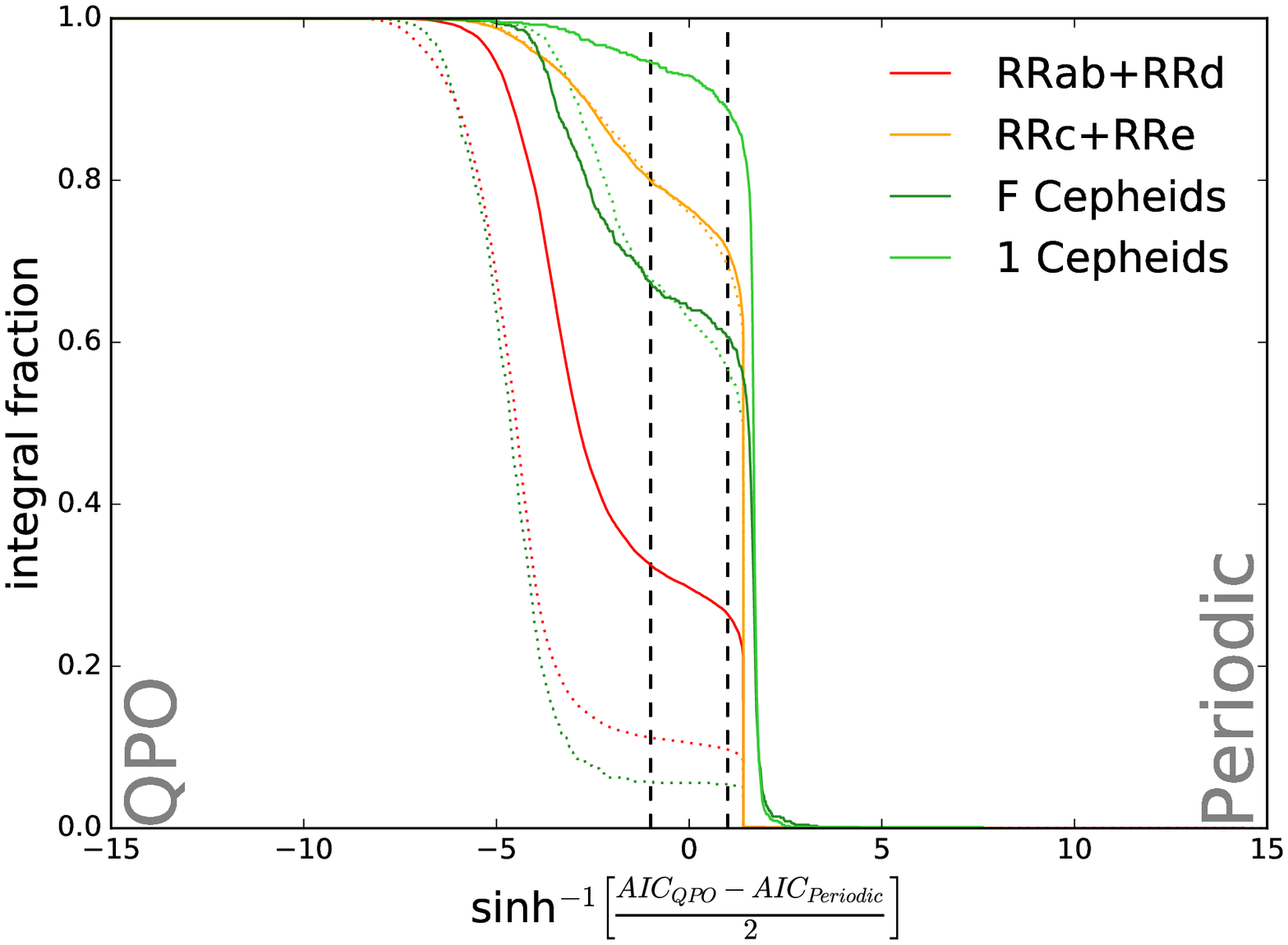}
  } 
\caption{Distribution of the RRab+RRd, RRc+RRe,
  fundamental Cepheids, and first overtone Cepheids in the AIC
  likelihood difference for the QPO and periodic models when using
  either the standard QPO model (dotted) or the higher order model
  (solid). The changes are small for the more sinusoidal RRc+RRe and
  first overtone Cepheids, but shift strongly toward the periodic
  case for the more sawtooth-like RRab+RRd and fundamental mode Cepheids.
  } 
\label{fig:p_v_quasi_p_2comp} 
\end{figure*} 

\subsection{Periodic versus Quasi-periodic} 
\label{sec:p_v_qpo}
We next test whether sources are better fit as truly periodic (i.e., $\tau \rightarrow \infty$~days) or as  
quasi-periodic (i.e., finite $\tau$) variables by comparing the likelihoods for a QPO with 
fixed $\tau/P = 10^4$ (as a proxy for $\tau \rightarrow \infty$~days) to those  with $\tau$ free to vary, 
as shown in Figure~\ref{fig:p_v_quasi_p}.  Quasars, as we would expect, strongly favour 
the QPO model, with 97 per cent preferring ($x<0$) the QPO model. Among the few exceptions, 
two have extreme outliers in the light curves (up to 25$\sigma$) and one has an exceptionally 
short light curve of 74 epochs.  As discussed earlier, when quasars are modeled as QPOs
they strongly favour either the ``incoherent'' limit ($\tau/P < 1$) 
where the QPO model becomes increasingly similar to the DRW model or
the white noise limit where $\tau$ is smaller than the cadence.

It is not surprising that the LPVs shown in
Figure~\ref{fig:p_v_quasi_p} generally favour a QPO over a periodic
model, as these variables generally do not have simple, coherent light
curve structures. Only 25 per cent of the Miras strongly prefer ($x>1$) a
periodic model, and most of the Miras that prefer a periodic model
have genuinely regular, sinusoidal light curves with mildly variable
($\sim 10$ per cent) amplitudes. The less regular SRVs show a smaller
periodic fraction while the low-amplitude OSARGs show an overwhelming preference for
the QPO model. The low amplitudes of the OSARGs also lead to noisier
light curves, where it becomes difficult to distinguish the models
(see Figure~\ref{fig:sawtooth} and the discussion in
\S\ref{sec:understanding}). The LPVs also have relatively low
coherence ratios, with $\sim 20$ per cent of Miras and less than $1$
per cent of SRVs and OSARGs having $\tau/P > 10$.

The surprise, leading to much of the discussion in \S\ref{sec:understanding},
is that a large fraction of periodic RR Lyrae and Cepheid variables
apparently favour a finite coherence time. For Cepheids, more than 30
per cent and 90 per cent of first overtone and fundamental
Cepheids, respectively, strongly 
prefer the QPO model even though they are very periodic
oscillators. RR Lyrae show a similar trend, with 20 per cent and 85
per cent of RRc
and RRab, respectively, preferring a QPO. This is not due to the presence of Blazhko
effect RR Lyrae alone, although such Blazhko RR Lyrae do tend to prefer the QPO
model (see below). For those objects that do prefer the QPO model, $95$ per cent of RRc
have coherence ratios $\tau/P > 100$, whereas $95$ per cent of RRab have
$\tau/P < 100$. Cepheids show a similar pattern, with more than 80 per cent
of fundamental mode Cepheids favouring a coherence ratio $\tau/P < 10$, while more than
$95$ per cent of first overtone Cepheids have $\tau/P > 10$. As
discussed in \S\ref{sec:understanding}, these behaviors
are a consequence of light curve shape. First overtone Cepheids, RRc, and
RRe variables all have relatively sinusoidal light curves, while fundamental mode
Cepheids, RRab, and RRd variables have more triangular/sawtooth
waveforms. Qualitatively, the integral fraction curves in Figure~\ref{fig:p_v_quasi_p} 
favour periodic models more strongly when the light curve is sinusoidal
and favour QPO models more strongly when the light curves are
sawtooth-like.

We introduced the higher order $p=4$ QPO models with periods locked in a 2:1 ratio and a common coherence time to explore if this difference
was driven by the assumption of sinusoidal waveforms in the one-component QPO
model. We show in Figure~\ref{fig:p_v_quasi_p_2comp} that using the two-component QPO models
halves the number of RRab and RRd variables that prefer a QPO compared to
the one-component model. The increase in periodic preference going from
a one-component to the two-component QPO models for the more sinusoidal
RRc and RRe sources is marginal. An even larger shift is seen for the
fundamental mode Cepheids. These results are in agreement with the
behavior we found in \S\ref{sec:understanding} for the experiments comparing artificial light
curves with sawtooth or sinusoid waveforms. It is not surprising,
therefore, that a higher-order QPO model results in a greater preference for periodicity. The two-component QPO is a better approximation to the sawtooth-like light
curves of RRab, RRd, and fundamental mode Cepheids, so the coherence
time can act more like the desired physical parameter, and less as a
proxy for light curve shape.

In addition to light curve shape, we also noted in \S\ref{sec:data} that
mischaracterized errors can bias the results, irrespective of
light curve shape. If some of the noise level assumed
in the model is less than the true noise, the extra variance will be
interpreted as signal. The natural way to model this extra variance is to
move the model away from periodic to model the mischaracterized noise
as stochastic variability. If we take the best-fitting model and
re-estimate the errors so that it has $\chi^2/dof = 1$, then the
fraction of Cepheids and RR Lyrae favouring the periodic model increase
significantly, from $70$ per cent and $10$ per cent to $90$ per cent
and $65$ per cent in the case of first overtone
 and fundamental Cepheids, respectively, and increasing from $80$ per cent and
 $15$ per cent among RRc/RRe and RRab/RRd to $95$ per cent and $90$
 per cent, respectively.

\begin{figure*}
\centering 
\includegraphics[width=0.9\textwidth]{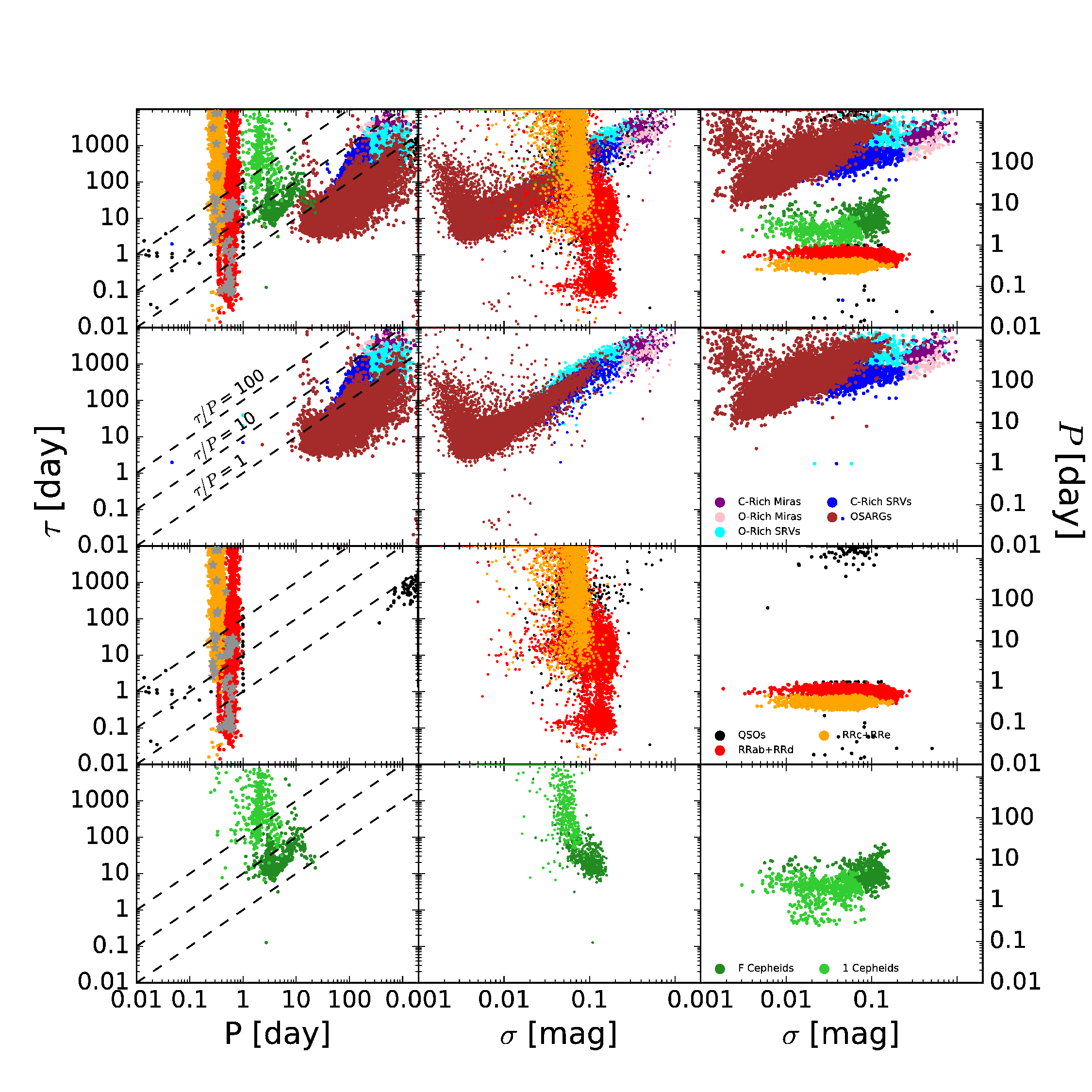} 
  \caption{Variable distributions in projections of the
    $P-\sigma-\tau$ QPO parameters. Note that
    chemical distinctions are included for SRVs and Miras. In the
    $P-\tau$ plane, dotted lines of constant coherence, $\tau/P$, are shown to
    guide the eye, where $\tau/P = 1$ is the boundary between
    coherent and incoherent periodicity. The first two columns show
    $\tau$ on the y-axis, whereas the third column is against
    $P$. Grey stars indicate RR Lyrae classified as Blazhko by \protect\cite{chen_jiang_yang2013}.}
\label{fig:all} 
\end{figure*}

\begin{figure*} 
\centering 
\includegraphics[width=0.5\textwidth]{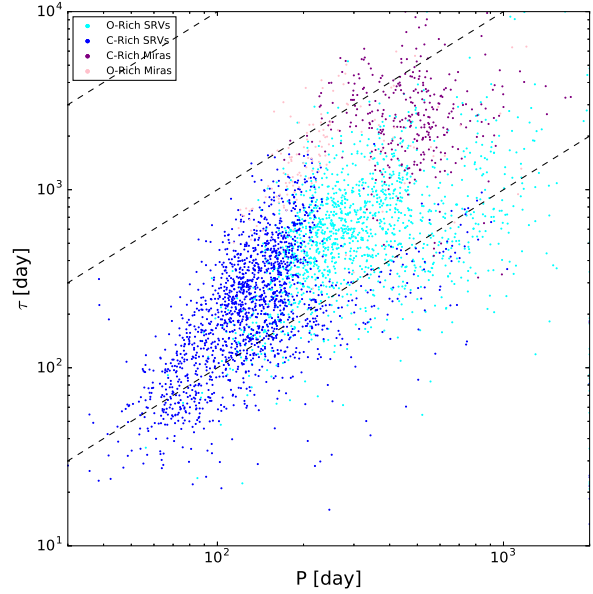}
\caption{A subset of the top left panel of Figure~\protect\ref{fig:all} showing only Miras and
  Semi-Regular Variables.} 
\label{fig:p_v_tau_lpvs} 
\end{figure*} 

\begin{figure*} 
\centering 
\includegraphics[width=0.5\textwidth]{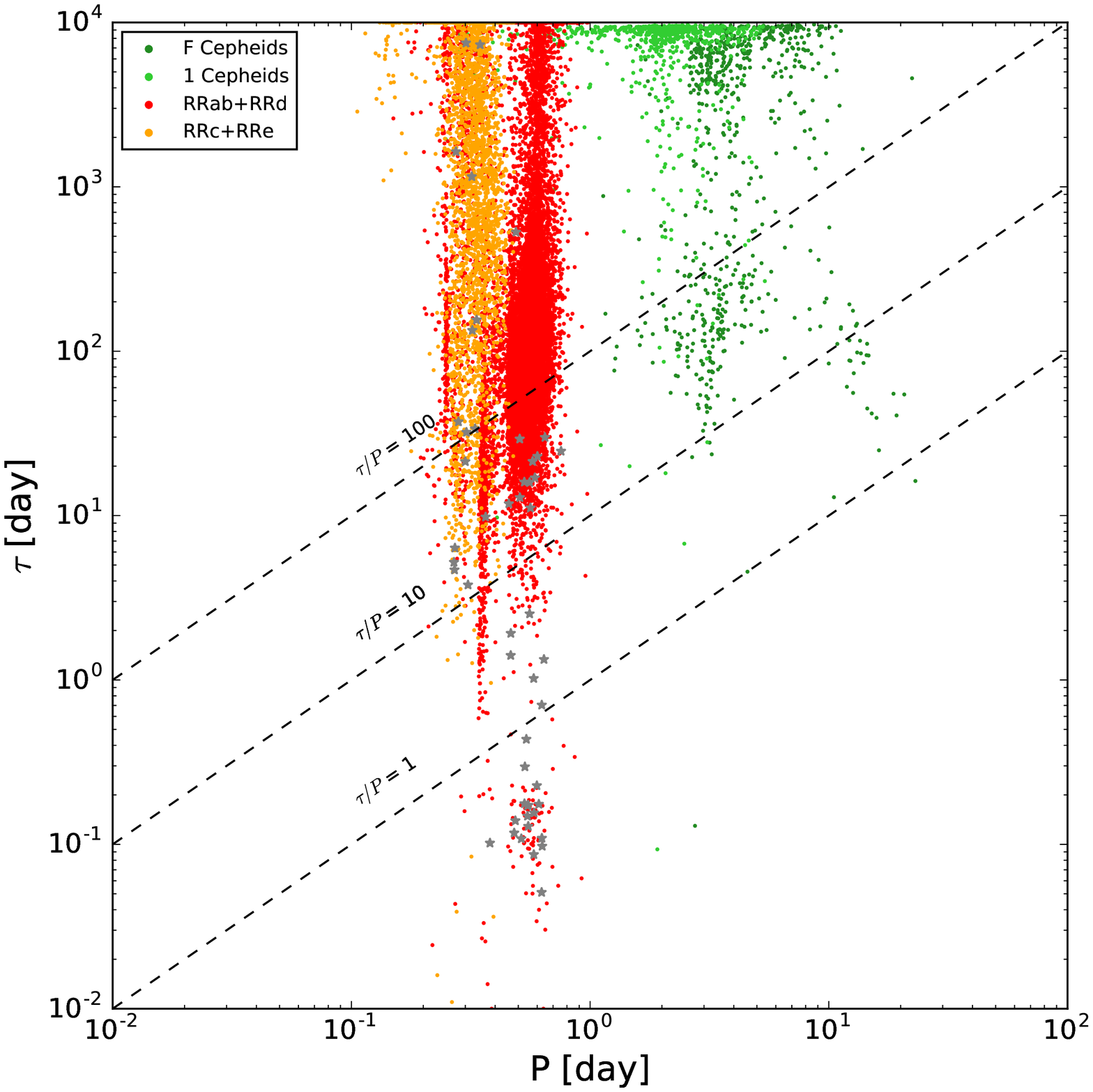}
\caption{The $P-\tau$ plane for RR Lyrae and Cepheids using the
  higher-order $p=4$ QPO model. Grey stars indicate RR Lyrae classified as Blazhko by \protect\cite{chen_jiang_yang2013}.} 
\label{fig:p_tau_2comp} 
\end{figure*}

\subsection{Variable Distributions in QPO Parameters } 
The lesson from \S\ref{sec:understanding} is that the QPO $\tau$ parameter
cannot simply be interpreted as a measure of coherence because it is
significantly affected by light curve shape. This is, however, no
barrier to using the QPO parameters to classify variables. Figure~\ref{fig:all} shows the distribution of all the sources in the $P-\tau$, $P-\sigma$, and $\tau-\sigma$ QPO parameter spaces. We see that the $P-\tau$ plane maps physically distinct variable types to nearly unique spaces. Below we briefly discuss the properties of each class of variable. We do not examine the properties of the variables in DRW parameter space since such an exploration was carried out in \cite{kozlowski+2010}.

If we first consider the distribution of the quasars, we see that they
largely have long periods ($P \ga 100$~days) and low
coherence ($\tau/P < 1$). There are also a small number with short
periods, small coherence time, and nominally higher coherence, as discussed in \S\ref{sec:qpo_v_drw} and Figure~\ref{fig:noise_qso}. QSOs overlap with part of the locus of the
OSARGs (discussed below), and therefore cuts in QPO space alone are
not sufficient to separate quasars from these stellar variables. However,
cuts in QPO/DRW model preference, as proposed in
\S\ref{sec:qpo_v_drw}, are efficient.

The various LPVs generally occupy somewhat different regions of the
QPO parameter space.  That they have differences in period and
amplitude is, of course, already known, so we focus on the the new
parameter $\tau$.  Figure~\ref{fig:p_v_tau_lpvs} shows the distribution of the Miras and SRVs
in $P$-$\tau$ space further divided by their C/O abundance class.
While not perfectly separated by $\tau$ at fixed period, it is
generally true that the Miras are more coherent (larger $\tau/P$) than
the SRVs
and that C-rich SRVs are more coherent than O-rich SRVs.  Similarly,
at fixed amplitude $\sigma$, the coherence time separates the O- and
C-rich SRVs well, and provides some discrimination between the O- and
C-rich Miras. It is probably not possible to well-separate the OSARGs
from the other LPVs simply based on their QPO parameters, possibly
because there are all simply part of an evolutionary continuum as
suggested by \cite{soszynski_wood_udalski2013}.  While our discussion in \S\ref{sec:understanding} and
the results of \S\ref{sec:results} suggest that interpreting $\tau$ simply
as a coherence time is risky, it is broadly true that longer period
Miras and SRVs have larger $\tau/P$ suggesting that they are more
coherent oscillators.  This is also seen simply from folded
light curves, where longer period Miras look more like the upper panels
of Figure~\ref{fig:bestfit} and shorter period ones look more like the lower panels.

Although they are highly regular variables, the Cepheids are not found
to have large coherence times $\tau$ in the QPO models for all the
reasons discussed earlier.  That $\tau$ appears primarily to be a
measure of light curve shape rather than coherence of oscillation,
does lead to a clear separation of the fundamental mode and overtone
Cepheids at fixed period.  Similarly, the RRc and RRe variables tend
to have larger $\tau$ than the RRab and RRd variables, but these
classes are so well separated in period that the QPO parameter adds
little new information.   Blazhko effect RR Lyra should appear as
low-coherence QPOs, although they are too rare to explain the overall
distributions.  If we look at the parameters of
the 52 confirmed Blazhko RR Lyrae from \cite{chen_jiang_yang2013} we see that
they are located preferentially at lower coherence ratios among RRab,
but are nearly uniform in distribution for RRc and RRe.  With a higher order model that would make all the
normal RR Lyrae coherent oscillators, the coherence time would likely
be an efficient means of identifying Blazhko variables. In particular,
Figure~\ref{fig:p_tau_2comp} shows the $P-\tau$ plane for the RR Lyrae and Cepheid variables
using the $p=4$ QPO model. Both RR Lyrae and Cepheids now have
significantly higher coherence times, as expected from the discussion
in \S\ref{sec:understanding}, and the Blazhko RR Lyrae are better separated.

\section{Discussion} 
\label{sec:conclusion} 
We have carried out a large scale survey of the properties of quasars
and variable stars using two different models of stochastic
variability.  The models are the two lowest order models within the
CARMA family of solutions to stochastic differential equations: (1)
the damped random walk (DRW) model characterized by an amplitude
$\sigma$ and a damping time $\tau$; and (2) the quasi-periodic
oscillation (QPO) model characterized by an amplitude $\sigma$, a
period $P$ and a coherence time $\tau$.  In general, variable stars
are better described by the QPO model while quasars are better
described by the DRW model. This is not a claim that the DRW is a
perfect model of QSO variability, although it is completely adequate
for our present purposes. There are some ambiguities in the details
of the model classification because the QPO model becomes
indistinguishable from the DRW model in the limits where the
oscillations are very incoherent ($\tau/P<1$), the period approaches
the time span of the data, or the coherence time, $\tau$, becomes
shorter than the observing cadence.  By combining the results from the two
stochastic models, we can significantly increase the efficiency in
separating quasars and variable stars over using the DRW model results
alone.

In principal, a quantity like the coherence time in the QPO models has
the potential to provide important physical information on the nature
of pulsations or, if applied to spotted stars, to the evolution time
scale of the spots. We found, however, that the estimated value of
$\tau$ in the QPO model is also quite sensitive to the deviations of
the light curve shape from a sinusoid. Qualitatively, this is most
easily understood in terms of the power spectrum of the variability.
The QPO model has a power spectrum which is a Lorentzian centred on
the principle frequency with a width set by the damping time scale.
For a variable with significant deviations from a sinusoid, such as
the ``sawtooth--like'' long period Cepheids or RRab variables, the
best $p=2$ QPO models have a finite coherence time in order to broaden the
Lorentzian to capture some of the variability power in the overtones of the
fundamental period.  If we use a higher order QPO model with
terms at both the fundamental period and the first overtone, in order
to better mimic the shape of such light curves, we find a longer
coherence time.   Thus, for the coherence time scale of these models
to have that physical meaning, the model must be of high enough order
to be able to match the shape of the underlying light curve.   This is
related to the models of a few individual variable sources by \cite{kelly+2014}, where they continued increasing the order of the stochastic
model until no increase in the order was statistically justified.  For
the RRab they considered, they
found that a higher-order CARMA$(7,0)$ model fit the light curve
best. \cite{kelly+2014} did not force any
fixed period ratios between the terms, but since many variables are
well-described by Fourier series \citep[e.g.,][]{simon&lee1981,simon&teays1982}, the approach we tested of
adding terms with fixed period ratios should work extremely well for
many variables.

While the QPO coherence time does not have a simple physical
interpretation because of these concerns, the QPO parameter space
still provides a means of separating variable classes by adding an
additional parameter beyond a period and an amplitude.   As noted
above, quasars tend to occupy regions of QPO parameter space distinct
from variable stars.  At fixed period, fundamental and first overtone
Cepheids have different coherence times, as do Oxygen- and Carbon-rich
Miras and SRVs.  While not a panacea for classification, models of
variable sources combining the results of DRW and QPO models provide
a relatively straight forward approach to providing quantitative
measures of light curves that can be used for classification.
Furthermore, the model fits can all be carried out in $O(N)$
operations using the algorithm discovered by \cite{ambikasaran+2014} even when evaluating
the full model likelihood rather than simply using the forecasting
approach of \cite{kelly+2014}.

Finally, the combination of the DRW and QPO models provides a
systematic approach to evaluating periodicity in quasars. There is
considerable interest in identifying periodic behavior in quasar
light curves as a probe of quasar binaries and there are now a number
of claimed detections \citep[e.g.,][]{sillanpaa+1988,maness+2004,rodriguez+2006,eracleous+2012,liu+2015,yan+2015,graham+2015,li+2016,zheng+2016,bhatta+2016}.  The problem is that standard means
of evaluating periodicity, such as a Lomb--Scargle periodogram \citep{lomb1976,scargle1982}, are essentially comparing the
probability of modeling the source as a sinusoid to the probability of
the same light curve being generated by a known level of white noise
characterized by the photometric errors.  As discussed by
\cite{press1978} and more recently by \cite{vaughan+2016}, this is
not a good statistical test when the underlying source class is known
to be variable and with a relatively red noise-like power spectrum
showing more variability on longer time scales.   We see this in our
analyses of the MQS quasars, where effects like noise levels or the
structure of the variability can make the QPO model of a quasar more
likely than the DRW model. The QPO model we consider here, which spans
variability from truly periodic back to the completely non-periodic
DRW model provides a good statistical basis for evaluating periodicity
in quasar light curves (modulo the caveats above) because it has a
continuous parameter $\tau$ to evaluate the significance of the
periodicity. Our approach also has the ability to systematically
remove light curve means or other calibration issues for light curves
combining multiple sources of data.

\section*{Acknowledgments}
CSK is supported
by NSF grant AST-1515927. We would like to thank Profs. M. Kubiak and G. Pietrzy{\'n}ski, former
members of the OGLE team, for their contribution to the collection
of the OGLE photometric data over the past years. The OGLE project has received funding from the Polish National Science
Centre grant MAESTRO no. 2014/14/A/ST9/00121 to AU. SK acknowledges
the financial support from the Polish National Science Center grant
no. 2014/15/B/ST9/00093.

\bibliographystyle{mnras}
\bibliography{bib_zinn_submit_141216}
 
\appendix 
 
\subsection{Trend Correction } 
 
The mean of the best-fitting light curves, $\hat{s}$, approximating
the observed light curve, $\vec{y}$, may be calculated using the
expression derived in \cite{rybicki_press1992},
\begin{equation}
\hat{s} = SC^{-1}(\vec{y} - L\vec{q}), 
\end{equation}
\noindent 
with variance from the true light curve given by 
\begin{equation}
\langle (\vec{s} - \hat{s})^2 \rangle = S - S^T C_{\perp} S, 
\end{equation}
\noindent 
where
\begin{equation}
C_{\perp} \equiv \left( C^{-1} -C^{-1}L(L^TC^{-1}L)^{-1}L^TC^{-1}\right)^{-1}. 
\end{equation}
\noindent 
As previously discussed in \cite{rybicki_press1992} and \cite{kozlowski+2010}, the CARMA formalism can be generalized to take into
account arbitrary trends. This assumes that the data are
modelled as
\begin{equation}
\vec{y} = \vec{s} + \vec{n} + L \vec{q}, 
\end{equation}
\noindent 
where $\vec{s}$ is the expected signal, described in this case by a
Gaussian process, $\vec{n}$ is a Gaussian noise term, and $L\vec{q}$
describes a trend. L is a $N\times \ell$ matrix, with $\ell$
representing the number of temporal trends to include in the fit and
$N$ is the number of data points. The column vector $\vec{q}$ is of
length $\ell$, and contains the polynomial coefficients for the
trend. In this notation, we can include a mean and linear trend in the
model, $q_1 + (t-t_0)q_2$, by setting $q = (q_1, q_2)$ and $L_{i1} = 1, L_{i2} = t_i - t_0$. 
 
As another example, to remove jumps within a light curve, like the
calibration shifts between the OGLE-III and IV QSO light curves, one
would again have $q = (q_1, q_2)$, but define $L$ to be $L =(1,0)$ for the M OGLE-III points of the light curve
and $L = (0,1)$ for the $N-M$ OGLE-IV points of the light curve.

%%%%%%%%%%%%%%%%%%%%%%%%%%%%%%%%%%%%%%%%%%%%%%%%%%

% Don't change these lines
\bsp% typesetting comment
\label{lastpage}
\end{document}